\documentclass[%
 reprint,
 amsmath,amssymb,
 aps,
]{revtex4-2}

\usepackage{graphicx}
\usepackage{dcolumn}
\usepackage{bm}
\usepackage{physics}
\usepackage{xr}
\usepackage{xcolor}
\usepackage[colorlinks=true,allcolors=black]{hyperref}
\usepackage{float}
\usepackage{placeins}

\newcommand{\startSIAppendix}{%
  \clearpage
  \onecolumngrid
  \setcounter{page}{1}%
  \renewcommand{\thepage}{S\arabic{page}}%
  \setcounter{section}{0}%
  \renewcommand{\thesection}{\Roman{section}}%
  \renewcommand{\thesubsection}{\thesection.\Alph{subsection}}%
  \setcounter{equation}{0}%
  \renewcommand{\theequation}{S\arabic{equation}}%
  \setcounter{figure}{0}%
  \renewcommand{\thefigure}{S\arabic{figure}}%
  \setcounter{table}{0}%
  \renewcommand{\thetable}{S\arabic{table}}%
  \begin{center}
    {\Large\bfseries SI Appendix}\\[1.5ex]
    {\bfseries Reaction-Coordinate-Dependent Non-Markovian Friction Governs Protein-Folding Dynamics}
  \end{center}
}

\begin{document}

\preprint{APS/123-QED}

\title{Reaction-Coordinate-Dependent Non-Markovian Friction Governs Protein-Folding Dynamics}

\author{Lucas Tepper}
\affiliation{Department of Physics, Freie Universit\"at Berlin, 14195 Berlin, Germany.}
\author{Benjamin J. A. Héry}
\affiliation{Department of Physics, Freie Universit\"at Berlin, 14195 Berlin, Germany.}
\author{Cihan Ayaz}
\affiliation{Department of Physics, Freie Universit\"at Berlin, 14195 Berlin, Germany.}
\author{Florian N. Brünig}
\affiliation{Department of Physics, Freie Universit\"at Berlin, 14195 Berlin, Germany.}
\author{Benjamin Dalton}
\affiliation{Department of Physics, Freie Universit\"at Berlin, 14195 Berlin, Germany.}
\author{Anton Klimek}
\affiliation{Department of Physics, Freie Universit\"at Berlin, 14195 Berlin, Germany.}
\author{Roland R. Netz}
\email{rnetz@physik.fu-berlin.de}
\affiliation{Department of Physics, Freie Universit\"at Berlin, 14195 Berlin, Germany.}

\date{\today}

\begin{abstract}
    \setlength{\parindent}{0pt}
 It is common to project the full atomic-resolution representation of a protein onto a
 one-dimensional reaction coordinate (RC) to capture the protein-folding kinetics. As a direct
 consequence of this dimensionality reduction, non-Markovian friction emerges
 in the framework of the generalized Langevin equation (GLE).
 All previous applications of GLEs to protein folding employed
 an RC-independent friction memory function and therefore did not account for the different friction
 in the folded and unfolded states. Using a recently derived GLE with
 RC-dependent mass and friction memory function, we introduce a novel method to extract
 memory functions from time series data via a conditional Volterra equation. When applied to molecular
 dynamics (MD) data of six fast-folding proteins, we find strongly RC-dependent
 memory friction in line with the intuitive expectation that friction is higher in
 the folded than in the unfolded state due to internal protein friction. Our numerically efficient method to
 simulate the GLE confirms the accuracy of the GLE parameter extraction by comparison with
 the MD data. We show that RC-dependent
 memory friction not only adds physical insight into the folding process but also significantly
 improves the description of protein folding kinetics using low-dimensional RCs.
\end{abstract}

\maketitle

Many biological processes involve proteins
in their native folded state, which requires the folding of a disordered polypeptide coil into a
well-defined, functional structure. Understanding the dynamics of this folding process is a
long-standing goal~\cite{dillProteinFoldingProblem502012}. Despite the high dimensionality of the protein conformational space, the folding process is often described using a
one-dimensional reaction coordinate (RC), which, for suitable RCs, captures the essential features of the process.
~\cite{choStructuralReactionCoordinates2006, bestReactionCoordinatesRates2005}.
Although several kinetic
theories have been used to describe the folding process~\cite{portmanMicroscopicTheoryProtein2001,travassoProteinFoldingTransition2010,
kayaConsistentModelingProtein2002, plotkinNonMarkovianConfigurationalDiffusion1998,
singhGeneralizedLangevinEquation2021, satijaGeneralizedLangevinEquation2019, frauenfelderEnergyLandscapesMotions1991,
zwanzigSimpleModelProtein1995, goTheoreticalStudiesProtein1983},
models based on the Langevin equation~\cite{
desanchoMolecularOriginsInternal2014, lickertModelingNonMarkovianData2020, bestDiffusiveModelProtein2006, lapidusEffectsChainStiffness2002}
have received particular attention.
These approaches describe the motion along the RC as a stochastic process governed by a free-energy landscape and frictional forces.
A central difficulty in such models is the correct treatment of the friction, which is not readily available experimentally~\cite{fosterProbingPositionDependentDiffusion2018}.
Many studies therefore investigate folding kinetics under varying solvent viscosities, extracting
solvent contributions and extrapolating to estimate internal protein friction~\cite{cellmerMeasuringInternalFriction2008,
zhengDependenceInternalFriction2015, schulzPeptideChainDynamics2012}.
An additional complication arises from the fact that friction is generally coordinate-dependent,
reflecting changes in the mobility along the folding pathway~\cite{bestCoordinatedependentDiffusionProtein2010,hinczewskiHowDiffusivityProfile2010,borgiaLocalizingInternalFriction2012}.
Even more challenging, it has become increasingly clear that non-Markovian effects play
an important role in protein dynamics~\cite{suarezImprovedEstimationLongTime2016,caoIntegrativeGeneralizedMaster2023,ayazNonMarkovianModelingProtein2021,daltonFastProteinFolding2023,sartoreMarkovTypeStateModels2025,klimekHierarchicalFrictionMemory2026}.
Memory effects naturally arise from the coupling between the
chosen RC and the many unresolved degrees of freedom of the full system.
These effects are captured within the framework of the generalized Langevin equation (GLE), which has proven valuable not only
for analyzing and simulating protein folding, but also for applications such as interpreting spectroscopic
data~\cite{brunigTimeDependentFrictionEffects2022}, studying cell
movement~\cite{mitterwallnerNonMarkovianDatadrivenModeling2020}, and predicting time-series
data~\cite{willersEfficientBayesianEstimation2024,kieferPredictionWeatherFinancial2025}.
In the context of protein folding, GLE-based approaches have demonstrated that neglecting memory effects
can lead to spurious RC-dependent friction profiles,
where apparent variations in friction arise not from intrinsic spatial heterogeneity,
but from projecting inherently non-Markovian dynamics onto a Markovian model~\cite{ayazNonMarkovianModelingProtein2021}.
In fact, a systematic treatment of RC-dependent friction effects within the non-Markovian GLE framework
is still lacking but essential for constructing reduced models that faithfully represent folding kinetics
and for interpreting experimentally inferred friction landscapes. \\
Many works~\cite{brunigTimeDependentFrictionEffects2022, ayazNonMarkovianModelingProtein2021,
daltonFastProteinFolding2023, darveComputingGeneralizedLangevin2009,
singhConformationalTransitionsAmyloidv2021, tepperAccurateMemoryKernel2024} used the heuristic GLE of the
form
\begin{figure*}
    \centering
    \includegraphics[trim={0 0.3cm 0 0.25cm},clip]{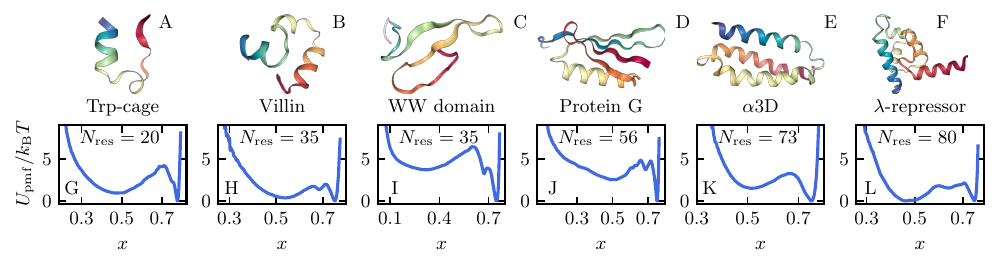} 
    \caption{ \textbf{A-F} The folded states of six fast-folding proteins taken from the PDB
 databank~\cite{bermanProteinDataBank2000}. \textbf{G-L} Corresponding potential landscapes
 $U_\mathrm{pmf}(x)$ for the fraction of native contacts RC $x$. All proteins show two-state folding
 dynamics between the unfolded state in the range of $0.3 < x < 0.55$, and the folded state around
 $x=0.75$. The proteins are ordered by their number of residues, $N_{\mathrm{res}}$, and the barrier
 height ranges from $2 \: k_{\mathrm{B}}T$ to $7 \: k_{\mathrm{B}}T$. } %
        \label{fig:intro_figure}%
\end{figure*}
\vspace{-0.2cm}
\begin{equation}
    \dv{t} v(t) = - \frac{1}{m_0} U_{\mathrm{pmf}}'(x(t)) - \int_{0}^{t} \dd s \: \Gamma(t - s) v(s) + F_R(t),
    \label{eq:GLE}
\end{equation}
in terms of the RC $x(t)$, the corresponding velocity $v(t) = \dd x(t) / \dd t$, the effective mass $m_0$,
the force due to the potential-of-mean-force $-U_{\mathrm{pmf}}'(x) = -\dd U_{\mathrm{pmf}}(x) / \dd x$, the memory
kernel $\Gamma(t)$ and the orthogonal force $F_R(t)$.
This GLE is inspired by pioneering works by
Mori~\cite{moriTransportCollectiveMotion1965} and
Zwanzig~\cite{zwanzigMemoryEffectsIrreversible1961,zwanzigNonlinearGeneralizedLangevin1973},
who started from a deterministic Hamiltonian in terms of atomic positions and momenta and projected the
dynamics onto a one-dimensional RC~\cite{vroylandtDerivationGeneralizedLangevin2022}.
Eq.~\ref{eq:GLE} combines an RC-independent
kernel from Mori's projection with the non-harmonic potential from Zwanzig's projection,
but in fact is approximate for general observables~\cite{vroylandtDerivationGeneralizedLangevin2022,heryGeneralizedLangevinEquations}.
Note that Mori's original GLE only features a harmonic potential, while Zwanzig's original GLE has a complex kernel depending on
the RC and its velocity, complicating the parametrization from time-series
data. Vroyland et
al.~\cite{vroylandtPositiondependentMemoryKernel2022} and Ayaz et
al.~\cite{ayazGeneralizedLangevinEquation2022} considered alternative projection operators that allow for the
systematic derivation of GLEs
including
RC-dependent friction functions.
In fact, Eq.~\ref{eq:GLE} follows from a systematically derived GLE~\cite{heryGeneralizedLangevinEquations}
involving an RC-dependent mass and a friction kernel $\Gamma(t)$ that
only depends on time
\begin{equation}
 \dv{t} v(t) = -\frac{ U_{\mathrm{eff}}'(x(t), v(t))}{m(x(t))}
        - \int_{0}^{t} \dd s \: \Gamma\left(t - s\right) v(s) + F_R(t),
    \label{eq:gam_of_x_GLE}
\end{equation}
under the condition that the mass $m(x)$ (defined below) has no RC-dependence, i.e., $m(x) = m_0$.
Another GLE of particular usefulness for protein folding takes the form~\cite{ayazSelfconsistentMarkovianEmbedding2022}
\begin{equation}
    \begin{split}
 \dv{t} v(t) &= -\frac{ U_{\mathrm{eff}}'(x(t), v(t))}{m(x(t))} \\
        &- \int_{0}^{t} \dd s \: \Gamma\left(t - s, x(s)\right) v(s) + F_R(t),
    \end{split}
    \label{eq:nonlinGLE}
\end{equation}
which differs from Eq.~\ref{eq:gam_of_x_GLE} by the RC-dependence of the kernel $\Gamma(t,
x)$. The effective potential
\begin{equation}
 U_{\mathrm{eff}}(x, v) = U_{\mathrm{pmf}}(x) + \frac{m(x)}{2} v^2 + \frac{k_{\mathrm{B}}T}{2} \ln m(x).
    \label{eq:ueff}
\end{equation}
contains the effective RC kinetic energy and a logarithmic mass term, which are important for
producing the correct stationary distribution as a function of $x$ and $v$, as shown below. Note
that the GLEs in Eqs.~\ref{eq:gam_of_x_GLE} and~\ref{eq:nonlinGLE} are exact, while the GLE in
Eq.~\ref{eq:GLE} only holds for $m(x) = m_0$, which is not a good approximation for proteins. \\
In this work, we establish how to extract the coordinate-dependent memory kernel in
Eq.~\ref{eq:nonlinGLE} directly from time-series data using a conditional Volterra equation applied
to molecular dynamics (MD) trajectories of six fast-folding proteins. Importantly, the GLE in
Eq.~\ref{eq:nonlinGLE} provides an exact and internally consistent coarse-grained description for
RC-dependent non-Markovian friction, as it systematically accounts for both memory effects and
spatial heterogeneities arising from the projection of the high-dimensional dynamics onto a
low-dimensional RC. We demonstrate that, for the considered proteins, the friction $\Gamma(t, x)$
depends strongly on $x$, exhibiting a pronounced maximum in the folded states. This finding
validates previous works highlighting the importance of internal friction in protein folding, which
predicted enhanced friction in compact, folded conformations compared to unfolded
ones~\cite{eatonModernKineticsMechanism2021, desanchoMolecularOriginsInternal2014,
borgiaLocalizingInternalFriction2012, alexander-katzInternalFrictionNonequilibrium2009}. By
introducing novel GLE simulation analyses and methodologies, we show that the RC-dependent memory
kernel yields crucial insights into the folding mechanism while also improving the accuracy
of the description of protein folding using GLEs. Notably, for proteins in which the position
dependence of the friction is particularly pronounced, RC-dependent GLE simulations clearly
outperform RC-independent models, thereby validating our approach and highlighting the relevance
of RC-dependent memory friction in protein-folding dynamics.

\begin{figure*}
    \centering
    \includegraphics[trim={0cm 0.17cm 0.0 0.25cm},clip]{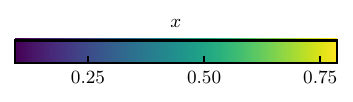}
    \includegraphics[trim={0 0 0 0.1cm},clip]{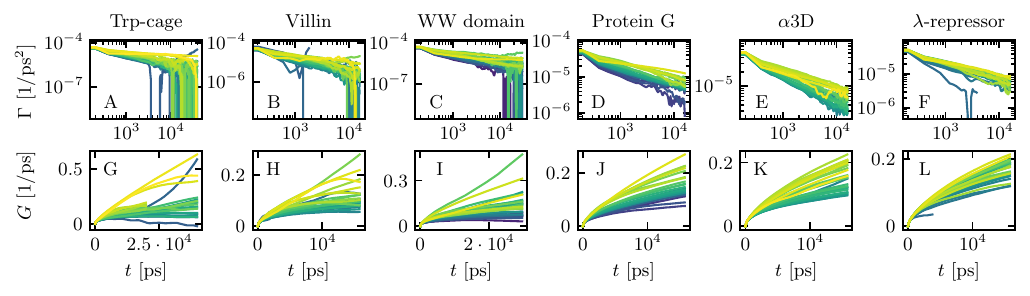}
    \includegraphics[trim={0cm 0.3cm 0.2cm 0.22cm},clip]{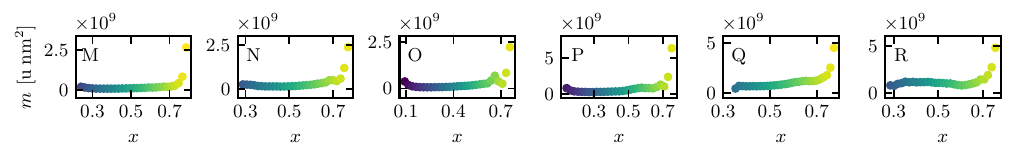}
    \caption{
        \textbf{A-F}
 RC-dependent kernels $\Gamma(t, x)$ and \textbf{G-L} integrated kernels $G(t,
 x)$, extracted from the Volterra Eq.~\ref{eq:derive_final} based on the GLE in Eq.~\ref{eq:nonlinGLE}. The color of the lines indicates
 the RC value $x$ (blue unfolded and yellow folded state).
        \textbf{M-R} RC-dependent masses $m(x)$, computed using Eq.~\ref{eq:mass_calc}. The units of $m(x)$
        follow from Eq.~\ref{eq:mass_calc} and the definition of the dimensionless RC
        in Eq.~\ref{eq:rc}.
 }%
    \label{fig:kernels}%
\end{figure*}
\section{Results}
We investigate RC-dependent non-Markovian effects using six different proteins: the Chicken
Villin subdomain (PDB 2F4K), the $\lambda$-D14A variant of the
$\lambda$-repressor~\cite{yangFoldingSpeedLimit2003} (PDB 1LMB), $\alpha_3D$ (PDB 2A3D), the WW
domain of the hPin1 cell cycle regulatory proline
cis/trans-isomerase~\cite{jagerStructureFunctionFolding2006}, the Trp-cage
protein~\cite{baruaTrpcageOptimizingStability2008} and the N37A/A46D/D47A mutant of the NuG2
protein~\cite{horngRapidCooperativeTwostate2003} (Protein G, PDB 1MIO). We base our analysis on previously published,
extensive all-atom molecular dynamics (MD) simulation data using the Anton supercomputer with
trajectory lengths from $125\:\mu\mathrm{s}$ for Villin up to $1154\:\mu\mathrm{s}$ for
Protein G~\cite{lindorff-larsenHowFastFoldingProteins2011}. To measure the folding process, we
employ a previously introduced RC based on the fraction of native contacts in the folded
state~\cite{bestNativeContactsDetermine2013, lindorff-larsenHowFastFoldingProteins2011}
\begin{equation}
    x(t) = \frac{1}{N} \sum_{i < j} \frac{1}{1 + e^{\beta ( s_{ij}(t) - \gamma s_{ij}^0 ) }},
    \label{eq:rc}
\end{equation}
where $N$ is the number of native-state contacts,  $s_{ij}(t)$ is the Cartesian distance between the
$C_{\alpha}$-atoms of residues $i$ and $j$ at time $t$, while $s_{ij}^0$ is the distance of the same
contact in the native state. We set the parameters $\beta = 30$ $\mathrm{nm}^{-1}$ and $\gamma = 1.6$~\cite{daltonFastProteinFolding2023}.
Figs.~\ref{fig:intro_figure} A-F depict the folded states of all six proteins, while
Figs.~\ref{fig:intro_figure} G-L show the potential of mean force $U_{\mathrm{pmf}}(x)$
derived from the equilibrium distribution $P_x(x)$ via $U_{\text{pmf}}(x) = -k_{\mathrm{B}}T \ln
P_x(x)$. For all proteins, $U_{\text{pmf}}(x)$ exhibits two distinct minima at the folded state at
approximately $x = 0.75$ and at the
unfolded state between $x = 0.3$ and $x = 0.55$, separated by a barrier of $2 \: k_{\mathrm{B}}T$ to $7 \: k_{\mathrm{B}}T$
(see SI Appendix for details). \\
To derive an extraction method for RC-dependent memory kernels, we start from the
GLE in Eq.~\ref{eq:nonlinGLE}
\begin{equation}
    \dv{t} v(t) = F(t)
 - \int_{0}^{t} \dd s \: \Gamma\left(t - s, x(s)\right) v(s) + F_R(t),
    \label{eq:derive1}
\end{equation}
where we have defined $F(t) = -U_{\mathrm{eff}}'(x(t), v(t)) / m(x(t))$. We multiply by $v(0)
\delta(x(0) - x)$, thereby conditioning $x(t)$ to start at the position $x$ at time $t=0$, and take
a phase-space average. Using the fact that the initial velocity is orthogonal to $F_R(t)$, i.e.,
$\langle v(0) \delta \left( x(0) - x \right) F_R(t) \rangle =
0$~\cite{heryGeneralizedLangevinEquations}, we obtain from Eq.~\ref{eq:derive1}
\begin{equation}
\begin{split}
    \dv{t} \langle v(0) \delta & \left(  x(0) - x \right) v(t) \rangle
 = \langle v(0) \delta \left( x(0) - x \right) F(t) \rangle \\
    & \: -\int_{0}^{t} \dd s \: \langle v(0) \delta \left( x(0) - x \right) \Gamma\left(t - s, x(s)\right) v(s) \rangle,
\end{split}
\label{eq:derive2}
\end{equation}
where $\langle A \rangle = \int_{\Omega} \dd \omega \:
 P_{\mathrm{eq}}(\omega) A(\omega)$ refers to the equilibrium average, $\omega$ to a
point in phase space $\Omega$ and $P_{\mathrm{eq}}(\omega)$ to the equilibrium canonical distribution.
We rearrange the integral in Eq.~\ref{eq:derive2} as $\int_{0}^{t} \dd s \: \langle v(0) \delta \left( x(0) - x \right) \Gamma\left(t - s, x(s)\right) v(s) \rangle
 = \int_{-\infty}^{\infty} \dd x_s \: \int_{0}^{t} \dd s \: C^{vv}(s, x, x_s) \Gamma\left(t - s, x_s\right)$,
where we have defined the double-conditional velocity autocorrelation function $C^{vv}(s, x,
x_s) = \langle v(0) \delta \left( x(0) - x \right) v(s) \delta\left( x(s) - x_s \right) \rangle$
with $x(0) = x$ and $x(s) = x_s$ at time $s$, and obtain
\begin{figure*}[t]
    \centering
    \includegraphics[trim={0 0.35 0 0.25cm},clip]{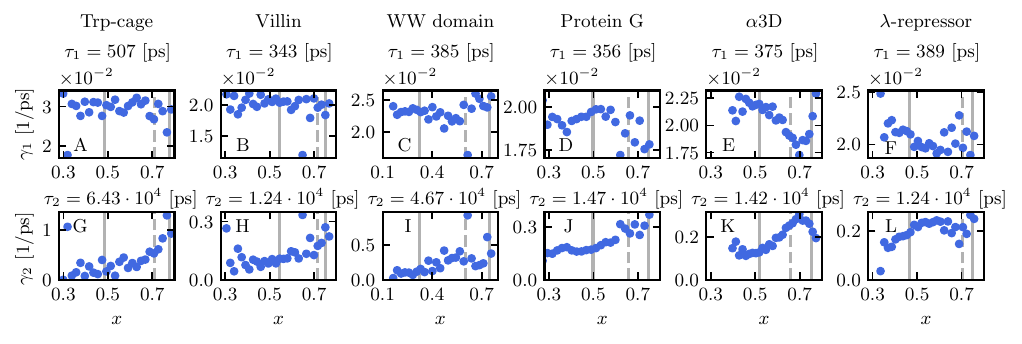}
    \caption{  Friction profiles $\gamma_i(x)$ for the two components of the multi-exponential fits following
 Eq.~\ref{eq:fit}. The subplot titles indicate the memory times $\tau_i$,
 Figs.~\textbf{A-F} show the fast memory component and Figs.~\textbf{G-L} the slower component.
 The vertical gray lines mark the minima of $U_{\text{pmf}}(x)$, the dashed gray line the location of the barrier.
 }%
    \label{fig:kernels1}%
\end{figure*}
\vspace{-0.03cm}
\begin{align}
    \label{eq:derive4}
        \dv{t} C^{vv}(t, x) &= C^{vF}(t, x) \\
 -&\int_{-\infty}^{\infty} \dd x_s \int_0^t \dd s \: C^{vv}(s, x, x_s) \Gamma(t - s, x_s). \nonumber
\end{align}
Here, we have used the single-conditional velocity autocorrelation function $C^{vv}(t, x) = \langle
v(0) \delta \left( x(0) - x \right) v(t) \rangle$ and velocity potential-gradient correlation
function $C^{vF}(t, x) = \langle v(0) \delta \left( x(0) - x \right) F(t)
\rangle$.
Due to numerical challenges in employing Eq.~\ref{eq:derive4} for memory kernel extraction, we
approximate $C^{vv}(s, x, x_s)$ as diagonal, i.e., $C^{vv}(s, x, x_s) = C^{vv}(s, x) \delta(x - x_s)$, yielding
\begin{equation}
    \dv{t} C^{vv}(t, x) = C^{vF}(t, x) - \int_0^t \dd s \: C^{vv}(s, x) \Gamma(t - s, x),
    \label{eq:derive5}
\end{equation}
the validity of this approximation will be checked below.
Direct computation of $\Gamma(t, x)$ from the Volterra
Eq.~\ref{eq:derive5} is possible~\cite{gordonGeneralizedLangevinModels2009,shinBrownianMotionMolecular2010}, but
tends to be unstable~\cite{langeCollectiveLangevinDynamics2006}. Instead, we extract the running integral
$G(t, x) = \int_0^t \dd s \:\Gamma(s, x)$ by integrating Eq.~\ref{eq:derive5}, leading to
\begin{equation}
    \begin{split}
        C^{vv}(t, x) - C^{vv}(0, x) = &\int_0^t \dd s \: C^{vF}(s, x) \\
        -&\int_0^t \dd s \: C^{vv}(s, x) G(t - s, x),
    \end{split}
    \label{eq:derive_final}
\end{equation}
which allows us to compute $G(t, x)$ from the conditional correlations $C^{vv}(t, x)$ and
$C^{vF}(t, x)$ via decoupled Volterra equations for each $x$ (details are given in
the SI Appendix). \\
Figs.~\ref{fig:kernels} A-F show the resulting kernels $\Gamma(t, x)$, and
Figs.~\ref{fig:kernels} G-L the running integrals $G(t, x)$. The memory extraction following
Eq.~\ref{eq:derive_final} is numerically stable up to $t_{\mathrm{max}}$ between 4 and $40 \:
\mathrm{ns}$ (see SI Appendix for more details). Figs.~\ref{fig:kernels} M-R show the
RC-dependent mass $m(x)$, defined as~\cite{ayazGeneralizedLangevinEquation2022}
\vspace{-0.25cm}
\begin{equation}
 m(x) = \frac{k_{\mathrm{B}}T}{\langle v(t)^2 \rangle_x} = \frac{k_{\mathrm{B}}T \langle \delta(x(t) - x) \rangle}{\langle \delta(x(t) - x) v(t)^2 \rangle}.
    \label{eq:mass_calc}
\end{equation}
We find that $m(x)$, $\Gamma(t, x)$, and $G(t, x)$ depend significantly on the RC value. The
mass increases towards the folded state (high $x$) for most proteins and rises sharply
in the folded basin. Likewise, the memory kernels increase significantly with rising $x$.
The higher friction in the folded state is expected since a compact protein has a higher internal
friction~\cite{eatonModernKineticsMechanism2021, desanchoMolecularOriginsInternal2014,
borgiaLocalizingInternalFriction2012, alexander-katzInternalFrictionNonequilibrium2009, einertConformationalDynamicsInternal2011a}. The mass
increase with $x$ is less intuitive and reflects non-linearities, as an RC linearly depending on atomic distances has a constant
mass~\cite{ayazGeneralizedLangevinEquation2022}. \\
To facilitate numerical simulations and analysis, we fit
the RC-dependent friction functions in Fig.~\ref{fig:kernels} to factorized multi-exponentials of the form
\begin{equation}
    \Gamma_{\mathrm{fit}}(t, x)=\sum_{i=1}^{N} \frac{\gamma_i(x)}{\tau_i} e^{-t / \tau_i}
    \label{eq:fit}
\end{equation}
with $N = 2$
(see SI Appendix for details).
Figs.~\ref{fig:kernels1} A-F present the friction profiles $\gamma_1(x)$ of the fast
memory component with memory times on the order of $300 \: \text{ps}$. Except for the $\alpha$3D
protein, all proteins exhibit a slight increase in $\gamma_1(x)$ approaching the folded state.
Figs.~\ref{fig:kernels1} G-L display the friction profiles $\gamma_2(x)$ of the slower memory
component, whose timescales range from $1.4 \: \text{ns}$ to $54 \: \text{ns}$, significantly longer
than those of $\gamma_1(x)$. The friction component $\gamma_2(x)$ exhibits a substantially larger
amplitude than $\gamma_1(x)$ and varies markedly with $x$, increasing toward the folded state near
$x = 0.75$. Given that the more slowly decaying friction component dominates the barrier-crossing
kinetics~\cite{lavacchiBarrierCrossingPresence2020}, Fig.~\ref{fig:kernels1} suggests that the
kinetic effect of the friction stems mostly from the folded state, where $\gamma_2(x)$ is
maximal.
\begin{figure*}
    \centering
    \includegraphics[trim={0.35cm 0 0 0.25cm},clip, scale=0.92]{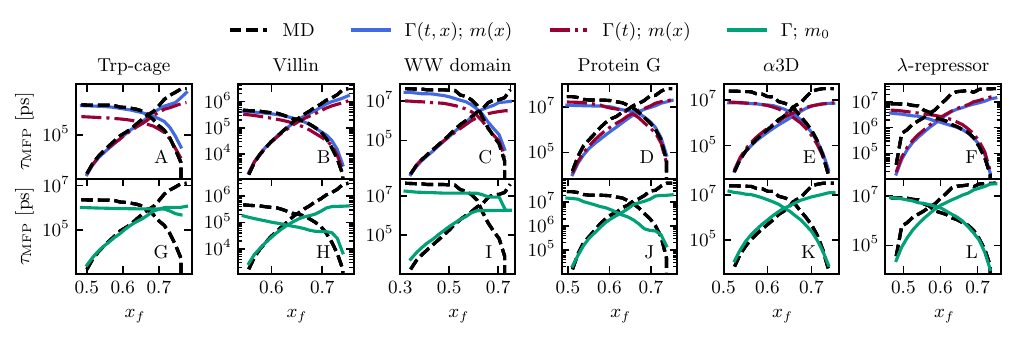}
    \caption{ Mean first-passage-time profiles $\tau_{\mathrm{MFP}}$ as a function of the final
 position $x_f$ of the MD data (black dashed line) compared to various GLE simulations for six
 fast-folding proteins for transitions starting from both minima in $U_{\mathrm{pmf}}(x)$ (see
 Fig.~\ref{fig:intro_figure}). Figs. \textbf{A-F} are based on kernel parameters
 extracted in this work. The blue lines show GLE simulations with RC dependence both in the
 kernel and the mass, using the exact GLE in Eq.~\ref{eq:nonlinGLE}. The red dashed lines show GLE simulations
 with an RC-dependent mass but a time-dependent kernel, using the exact GLE in
 Eq.~\ref{eq:gam_of_x_GLE}.
 The green lines in Figs. \textbf{G-L} show results from the approximate GLE in Eq.~\ref{eq:GLE},
 with RC-independent mass and kernel, based on kernel parameters from Ref.~\cite{daltonFastProteinFolding2023} and parametrized by
 Eq.~\ref{eq:kernel_ben}.
 }%
    \label{fig:mfpt}%
\end{figure*}

To check the validity of our estimated parameters, we simulate the GLE in
Eq.~\ref{eq:nonlinGLE} with a coordinate-dependent kernel using a Markovian embedding scheme based on the
equations of motion for the RC $x(t)$ and the auxiliary variables $y_i(t)$
\begin{align}
        \dot{x}(t) &= v(t), \nonumber \\
 \dot{v}(t) &= -\frac{1}{m(x(t))}\dv{x(t)} U_{\text{eff}}(x(t), v(t)) + \sum_i \frac{y_i(t)}{\tau_i}, \nonumber \\
        \dot{y}_i(t) &= -\frac{y_i(t)}{\tau_i} - \bar{\gamma}_i(x(t)) m(x(t)) v(t) + \frac{\bar{\gamma}_i'(x(t))}{2 \bar{\gamma}_i(x(t))} v(t) y_i(t) \nonumber \\
        &+\sqrt{k_{\mathrm{B}}T \bar{\gamma}_i(x(t))} \eta_i(t),
    \label{eq:eoms}
\end{align}
where $\eta_i(t)$ is Gaussian white noise, defined by $\langle \eta_i(t) \rangle = 0$, $\langle
\eta_i(t) \eta_j(t' )\rangle = 2 \delta_{ij} \delta(t - t')$ and $\bar{\gamma}_i(x) = \gamma_i(x) /
m(x)$. Equation~\ref{eq:eoms} reproduces the GLE in
Eq.~\ref{eq:nonlinGLE}, yields the stationary distribution $P(x, v) \propto
\sqrt{m(x)} P_x(x) \exp{-m(x) v^2 / (2 k_{\text{B}}T)}$ and the
marginalized positional distribution $P_x(x) \propto \exp{-U_{\text{pmf}}(x) / k_{\text{B}}T}$.
The SI Appendix
demonstrates that the residual differences between the kernel produced by Eq.~\ref{eq:eoms} and the
target form of Eq.~\ref{eq:fit} are negligible.
This is demonstrated in
Fig.~\ref{fig:mfpt}, where we present the mean first-passage times $\tau_{\text{MFP}}(x_s, x_f)$,
defined as the mean time to start from $x_s$ and reach $x_f$ for the first time,
for barrier-crossing processes originating either in the unfolded or folded state. The blue lines in Fig.~\ref{fig:mfpt} A-F
indicate the results of the embedding in Eq.~\ref{eq:eoms}, which includes $\Gamma(t, x)$ and $m(x)$,
and are in good agreement with
the MD reference kinetics (black dashed lines) for most proteins. To highlight the importance of RC-dependent
effects, we compare the embedding in Eq.~\ref{eq:eoms} to simulations based on the approximate GLE in
Eq.~\ref{eq:GLE}, without an RC-dependent mass or kernel. Here, we use the kernel
\vspace{-0.15cm}
\begin{equation}
    \Gamma_{\mathrm{fit}}(t) = \sum_{i=1}^{N} \frac{\gamma_i}{\tau_i} e^{-\frac{t}{\tau_i}},
    \label{eq:kernel_ben}
\end{equation}
based on parameters previously estimated from the MD data~\cite{daltonFastProteinFolding2023} (see
SI Appendix for more details and for a comparison with previous results).
$\tau_{\text{MFP}}$ from the GLE in Eq.~\ref{eq:GLE} (green lines in Figs.~\ref{fig:mfpt} G-L) match
the MD reference well for some proteins, like the $\lambda$-repressor, but fail for other proteins,
like Trp-cage. The non-Markovian simulations, based on the exact GLE in Eq.~\ref{eq:nonlinGLE} with
RC-dependent mass and friction, deliver notable improvements in $\tau_{\text{MFP}}$ for Villin, WW
domain and Trp-cage, for which the RC-dependence of the friction in Fig.~\ref{fig:kernels1} is
large, while being of comparable quality for the other proteins. The only exception is the
$\lambda$-repressor, which has a largely RC-independent friction profile and for which the
approximate GLE in Eq.~\ref{eq:GLE} performs slightly better than the exact GLE in
Eq.~\ref{eq:nonlinGLE}, which will be discussed below. Having demonstrated that the RC-dependence
of $m(x)$ and $\Gamma(t, x)$ plays a crucial role, we seek to separate the effects of $m(x)$ and
$\Gamma(t, x)$. For this, we use the exact GLE in Eq.~\ref{eq:gam_of_x_GLE} involving an
RC-dependent mass, but an RC-independent kernel. We extract the corresponding kernel from the MD
data using the coordinate-independent analog to the Volterra equation~\ref{eq:derive_final},
followed by a numerical fit as in Eq.~\ref{eq:kernel_ben}. The SI Appendix shows that the kernels obtained from the GLEs in Eqs.~\ref{eq:GLE}
and~\ref{eq:gam_of_x_GLE} are remarkably similar. The numerical GLE simulations of
Eq.~\ref{eq:gam_of_x_GLE} use the embedding in Eq.~\ref{eq:eoms}. Figs.~\ref{fig:mfpt} A-F show that
the mean first-passage times (red dashed lines) for GLE simulations using $\Gamma(t)$ and $m(x)$ are
in close agreement with the MD reference (black dashed lines). These results are similar, but
slightly inferior to simulation results using the GLE in Eq.~\ref{eq:nonlinGLE}, which includes
RC-dependent mass and friction. This is at first sight surprising, since both GLEs in
Eqs.~\ref{eq:gam_of_x_GLE} and~\ref{eq:nonlinGLE} are exact, and suggests that the GLE in
Eq.~\ref{eq:nonlinGLE} including the RC-dependent friction is numerically more robust. Another
possible source of discrepancy between simulations based on different GLEs is how accurately
non-Gaussian contributions to the orthogonal force, which presumably are present in MD data, are
reproduced. This may well explain the observed superior performance of GLE simulations based on
Eq.~\ref{eq:GLE} relative to Eq.~\ref{eq:nonlinGLE} when compared with the MD results for the $\lambda$-repressor in terms of cancellation of error.

Our results demonstrate that the RC-dependence of the friction memory function not only reveals
protein internal friction in agreement with previous experiments and simulations~\cite{chungStructuralOriginSlow2015, Best2010CoordinateDependentDiffusion, borgiaLocalizingInternalFriction2012, desanchoMolecularOriginsInternal2014}, but also improves the GLE-based description of protein-folding kinetics.
The approximation-free GLEs in Eqs.~\ref{eq:gam_of_x_GLE} and~\ref{eq:nonlinGLE} broaden the
applicability of non-Markovian theory to include such RC-dependent effects.

\section*{Discussion and Conclusion}
From an exact formulation of the GLE, including RC-dependent mass and friction kernel, we derive a
novel Volterra equation, allowing us to compute RC-dependent memory kernels from time series data in
an entirely data-driven analysis. The extracted kernels are validated by establishing a Markovian
embedding scheme that allows for the simulation of GLEs, which include RC-dependent masses and memory
kernels. Applying our framework to six fast-folding proteins, we find that for proteins that exhibit
pronounced RC-dependent friction, which is indicative of internal friction, simulations based on the
approximate generalized Langevin equation without RC-dependent mass and friction are insufficient to reproduce the MD
folding and unfolding kinetics. More broadly, our analysis identifies strong RC-dependence of the
memory kernel as a feature of many, but not all, proteins. For all proteins except
$\lambda$-repressor, the extracted kernels peak in the folded, compact state, where amino acids are
closely packed and interact strongly. These results emphasize that coordinate-dependent friction is
a fundamental feature of protein folding dynamics rather than a minor correction. Incorporating
RC-dependent, non-Markovian friction therefore provides both a more faithful physical picture and a
quantitatively improved low-dimensional description of protein dynamics. We expect that this
framework will extend naturally to larger proteins and other macromolecular systems, unveiling new
possibilities for understanding dynamical mechanisms in biomolecular processes.

\section*{Materials and Methods}
The computation of the RC and details about the MD data are given in SI Appendix, section I. The
numerical kernel extraction procedure is explained in SI Appendix, section II. Details about the
kernel fitting are given in SI Appendix, section III. The GLE simulation method is derived in SI
Appendix, section IV. Numerical aspects of the GLE simulations are given in SI Appendix, section V.
SI Appendix, section VI explains the kernel extraction using the GLE in Eq.~\ref{eq:gam_of_x_GLE}
and SI Appendix, section VII compares with previous results.

\section*{Data Availability}
The code used for the kernel extraction is available at
\href{https://github.com/lucastepper/correlation}{https://github.com/lucastepper/correlation} and
\href{https://github.com/lucastepper/kernel_extraction}{https://github.com/lucastepper/kernel\_extraction}.

\section*{Acknowledgments}
This work was funded by the Deutsche Forschungsgemeinschaft (DFG, German Research
Foundation) - ID 431232613 - SFB 1449 project A02.

\bibliography{main}

\startSIAppendix


\FloatBarrier
\section{\label{si:mdData}Molecular Dynamics Data}
%
We consider the folding dynamics of six fast-folding proteins, Chicken Villin subdomain (PDB 2F4K), the
$\lambda$-D14A variant of the $\lambda$-repressor~\cite{yangFoldingSpeedLimit2003} (PDB 1LMB),
$\alpha_3D$ (PDB 2A3D), the WW domain of the hPin1 cell cycle regulatory proline
cis/trans-isomerase~\cite{jagerStructureFunctionFolding2006}, the Trp-cage
protein~\cite{baruaTrpcageOptimizingStability2008} and the N37A/A46D/D47A mutant of the NuG2
protein~\cite{horngRapidCooperativeTwostate2003} (Protein G, PDB 1MIO). Our analysis is based on
previously published all-atom molecular dynamics (MD) simulation data obtained with the Anton supercomputer
with trajectory lengths starting from $125\:\mu\mathrm{s}$ for Villin up to $1154\:\mu\mathrm{s}$
for Protein G~\cite{lindorff-larsenHowFastFoldingProteins2011}. As in our previous
publication~\cite{daltonFastProteinFolding2023}, the fraction of native contact reaction coordinate
$x$ is computed from the $C_{\alpha}$-atoms of the MD trajectory, recorded at a 200-ps time step.
The computation of $x$ requires a reference state, for which we take the native state. To find the
native state, we follow the implementation used by Lindorff-Larsen et
al.~\cite{lindorff-larsenHowFastFoldingProteins2011} and Best et
al.~\cite{bestNativeContactsDetermine2013}. Briefly, for each frame of the trajectory, we count the
number of 'neighbouring' frames, defined as frames that have a root-mean-squared deviation less than
0.2 nm. We take the frame with the most 'neighbouring' frames as the folded state. In the native
state, we define all $C_{\alpha}$-pairs that are at least five residues apart in the primary
sequence and are separated by a distance of less than 0.9 nm as the native contacts. We compute the
fraction of native contact reaction coordinate via
\begin{equation}
 x(t) = \frac{1}{N} \sum_{i < j} \frac{1}{1 + e^{\beta ( s_{ij}(t) - \gamma s_{ij}^0 ) }},
\end{equation}
where $N$ is the number of native contacts,  $s_{ij}(t)$ is the Cartesian distance
between the $C_{\alpha}$-atoms of the native contact between residues $i$ and $j$ at time $t$, while
$s_{ij}^0$ is the same contact distance in the native state. We set the parameters $\beta = 30$
$\mathrm{nm}^{-1}$ and $\gamma = 1.6$~\cite{daltonFastProteinFolding2023}.

\FloatBarrier
\section{\label{si:numerics_volterra}Numerical implementation of the Conditional Volterra Equation}
In the main text, we derive the conditional Volterra equation
\begin{equation}
 C^{vv}(t, x_k) - C^{vv}(0, x_k) = \int_0^t \dd s \: C^{vF}(s, x_k)
 -\int_0^t \dd s \: C^{vv}(s, x_k) G(t - s, x_k).
    \label{eq:derive_final_main}
\end{equation}
To implement Eq.~\ref{eq:derive_final_main} numerically, we discretize all functions of time as
$f(t, x_k) = f(i \Delta t, x_k) = f_i(x_k)$ and discretize the integrals using the trapezoidal rule
\begin{equation}
    \begin{split}
 C^{vv}_i(x_k) - C^{vv}_0(x_k) = &\frac{\Delta t}{2} \left( C^{vF}_0(x_k) + C^{vF}_i(x_k)
 - C^{vv}_i(x_k) G_0(x_k) - C^{vv}_0(x_k) G_i(x_k) \right) \\
        &-\Delta t \sum_{j=1}^{i - 1} \left( C^{vv}_j(x_k) G_{i-j}(x_k) - C^{vF}_j(x_k) \right).
    \end{split}
    \label{eq:num1}
\end{equation}
Solving for $G_i(x_k)$ and inserting $G_0(x_k) = 0$, we obtain the iterative equation for given $x_k$
\begin{equation}
    \begin{split}
 G_{i}(x_k) =  &\frac{1}{C^{vv}_{0}(x_k)} \left(
 C^{v F}_{0}(x_k) + C^{v F}_{i}(x_k)
 + \frac{2}{\Delta t} \left(C^{vv}_{0}(x_k) - C^{vv}_{i}(x_k) \right)
 \right) \\
 -& \frac{2}{C^{vv}_{0}(x_k)} \sum_{j = 1}^{i - 1} \left( G_{i-j}(x_k) C^{vv}_{j}(x_k) - C^{vF}_j(x_k) \right).
    \end{split}
    \label{eq:num2}
\end{equation}
We compute the  time correlation functions $C^{ab}(t, x_k) = \langle a(0) \delta(x(0) - x_k) b(t)
\rangle$ via a discrete version of the Wiener-Khinchin
theorem~\cite{wienerGeneralizedHarmonicAnalysis1930,khintchineKorrelationstheorieStationaerenStochastischen1934}
\begin{equation}
    C^{ab}_i(x_k)=\frac{N}{(N-i) \sum_{j = 1}^N I_k[x_j]}\,
    IDFT\!\left[DFT\!\left(I_k[x_i]\odot a_i\right)\odot DFT[b_i]^*\right],
\end{equation}
where $a_i = a(i \Delta t)$ and $b_i = b(i \Delta t)$ are the two discrete time series to correlate and
$x_i = x(i \Delta t)$ is the position the correlation is conditioned on. The time series $a_i$ and $b_i$
are padded with zeros to twice their original length $N$ by appending zero entries to the time series.

\begin{equation}
 DFT[a_n] = \sum_{n=0}^{N-1} a_n \, e^{-2\pi j \frac{\omega n}{N}}, \quad \omega = 0, 1, \dots, N-1,
\end{equation}
is the discrete Fourier transform,
\begin{equation}
 IDFT[a_\omega]  = \frac{1}{N} \sum_{\omega=0}^{N-1} a_\omega \, e^{2\pi j \frac{\omega n}{N}}, \quad n = 0, 1, \dots, N-1,
\end{equation}
is the inverse discrete Fourier transform (implemented via FFT~\cite{cooleyAlgorithmMachineCalculation1965}),
where $N$ is the length of the time series $a_i$ and $j = \sqrt{-1}$.
To condition on the positions $x_i$, we multiply element-wise with the indicator function
\begin{equation}
I_k[\cdot] =
\begin{cases}
1, & \text{if } x_k - \frac{\Delta x}{2} \leq x < x_k + \frac{\Delta x}{2} \\
0, & \text{otherwise}
\end{cases}
\end{equation}
for bins at positions $x_k$ with width $\Delta x$. We denote element-wise multiplication by $\odot$
and complex-conjugation by $^*$. \\
\\
To calculate $U_{\text{pmf}}(x)$, we estimate $P(x)$ using a histogram with 200 bins and then use
$U_{\text{pmf}}(x) = -k_{\mathrm{B}}T \log P(x)$. Next, we compute $U'_{\text{pmf}}(x)$ via a numerical
gradient of $U_{\text{pmf}}(x)$. A time series of $U'_{\text{pmf}}(x(t))$ is obtained
by linearly interpolating $U'_{\text{pmf}}(x)$ at the MD simulation positions $x(t)$.

To compute
$U_{\text{eff}}(x, v)$ according to Eq.~\ref{eq:ueff} in the main text and convert it to a time series for the correlation functions, we start with the histogram of $U_{\text{pmf}}(x)$. Since it has 200 bins,
while only 30 bins are used to compute $m(x)$, we use a linear interpolation to compute
$\log(m(x))$ evaluated at the bins of the histogram of $U_{\text{pmf}}(x)$ and
obtain $U_{\text{pmf}}(x) + \frac{k_{\mathrm{B}}T}{2} \log(m(x))$. Next,
we linearly interpolate this expression at the MD simulation positions $x(t)$. For this,
we also linearly interpolate $m(x)$ at the MD simulation positions $x(t)$
to compute the kinetic energy $m(x(t)) \frac{v^2(t)}{2}$.
Finally, we add these two interpolated time series, yielding $U_{\text{eff}}(x(t), v(t))$.
\FloatBarrier
\section{\label{si:fitting}Fitting the Position-Dependent Kernels}
\begin{figure}
    \centering
    \includegraphics[trim={0 0.2cm 0 0},clip]{figures/shaw_cbar.pdf} 
    \includegraphics[trim={0 0 0 0.1cm},clip]{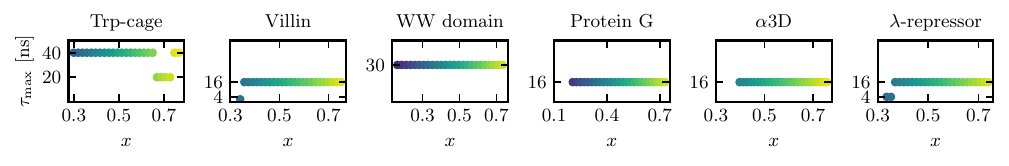} 
    \caption{We show the maximum time $\tau_{\mathrm{max}}(x_k)$ the Volterra-based extraction
 resolves during the kernel computation due to numerical accuracy in each bin $x_k$. We consider values
    $\tau_{\max}(x_k) \in \{4, 16, 20, 30, 40\} \: \mathrm{ns}$. }%
    \label{fig:shaw_ends}%
\end{figure}

\begin{figure}
    \centering
    \includegraphics[trim={0 0 0 0},clip]{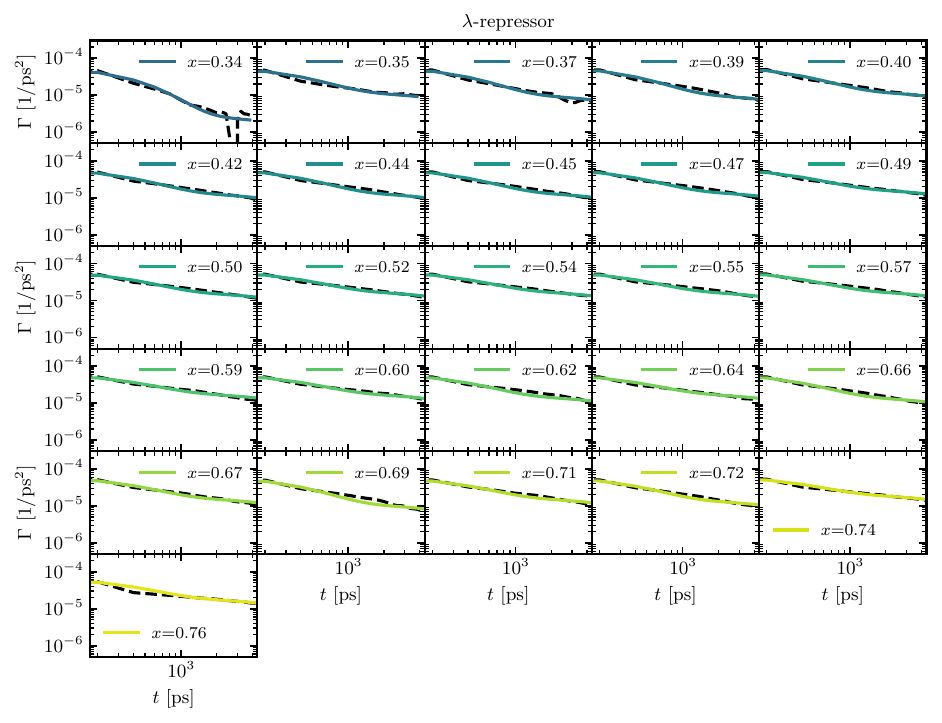} 
    \caption{
 We compare the numerically extracted kernels $\Gamma(t, x)$ (coloured lines) to the
 corresponding fits for Villin.
 }%
    \label{fig:fits_for_Villin}%
\end{figure}

\begin{figure}
    \centering
    \includegraphics[trim={0 0 0 0},clip]{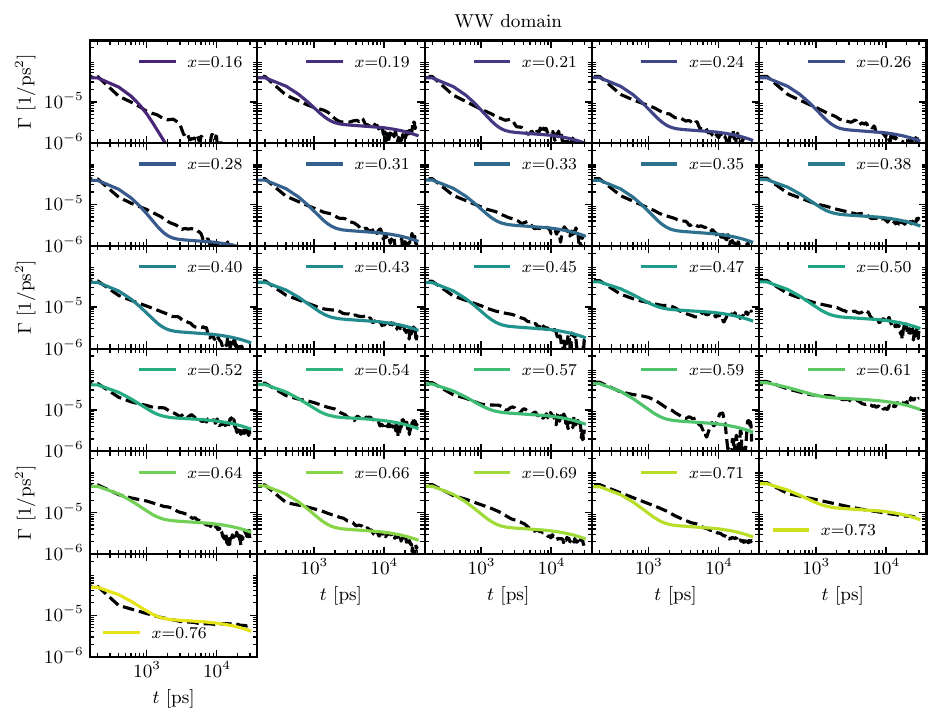} 
    \caption{
 We compare the numerically extracted kernels $\Gamma(t, x)$ (coloured lines) to the
 corresponding fits for $\lambda$-repressor.
 }%
    \label{fig:fits_for_lambda_repressor}%
\end{figure}

\begin{figure}
    \centering
    \includegraphics[trim={0 0 0 0},clip]{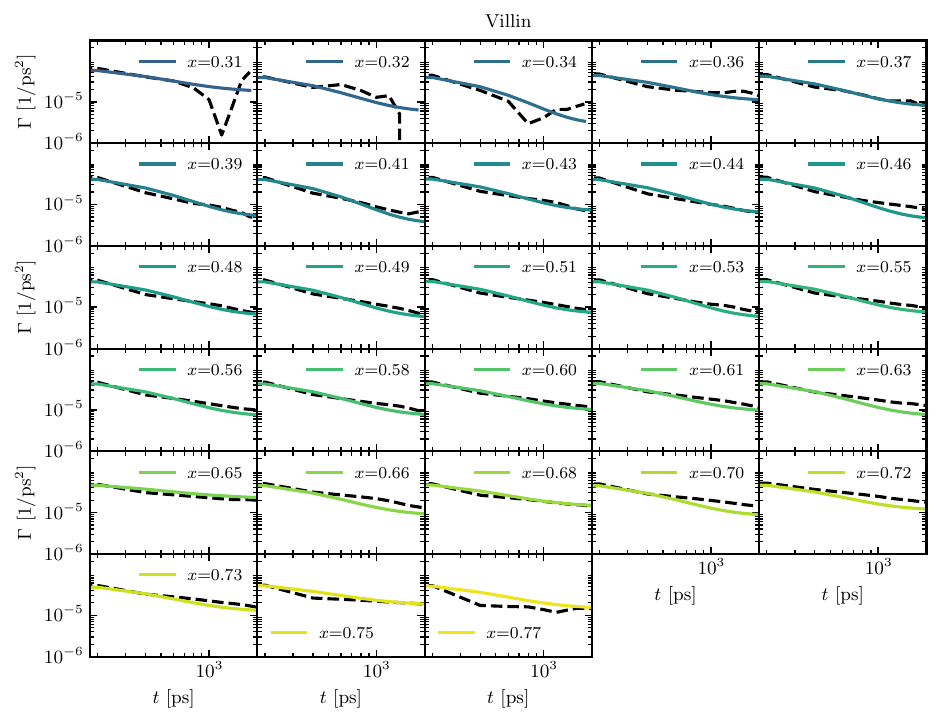} 
    \caption{
 We compare the numerically extracted kernels $\Gamma(t, x)$ (coloured lines) to the
 corresponding fits for Trp-cage.
 }%
        \label{fig:fits_for_TrpCage}%
    \end{figure}

\begin{figure}
    \centering
    \includegraphics[trim={0 0 0 0},clip]{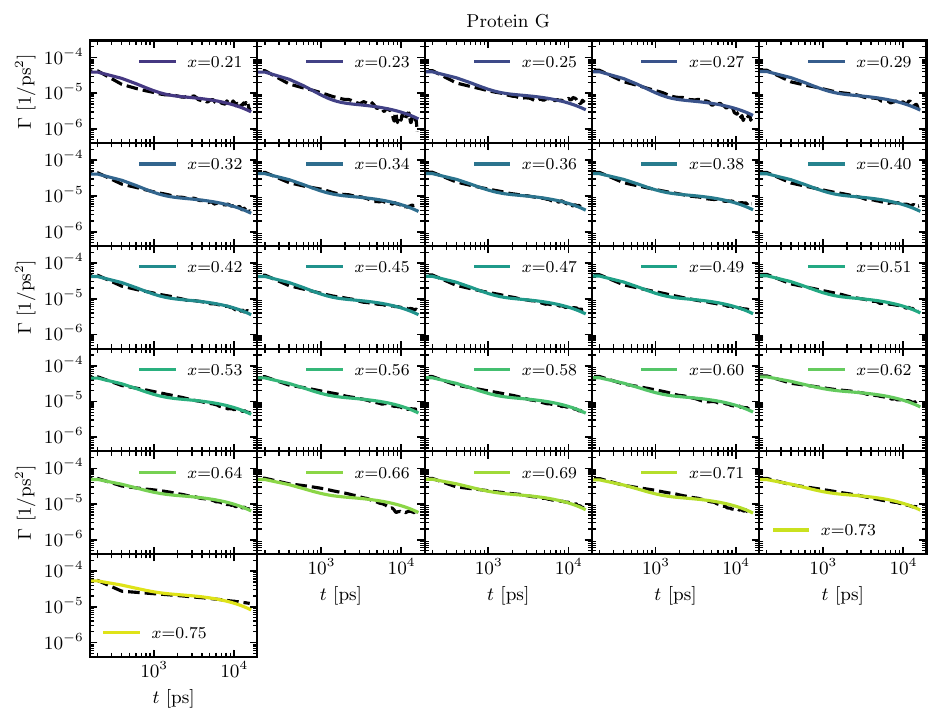} 
    \caption{
 We compare the numerically extracted kernels $\Gamma(t, x)$ (coloured lines) to the
 corresponding fits for $\alpha$3D.
 }%
    \label{fig:fits_for_alpha3D}%
\end{figure}

\begin{figure}
    \centering
    \includegraphics[trim={0 0 0 0},clip]{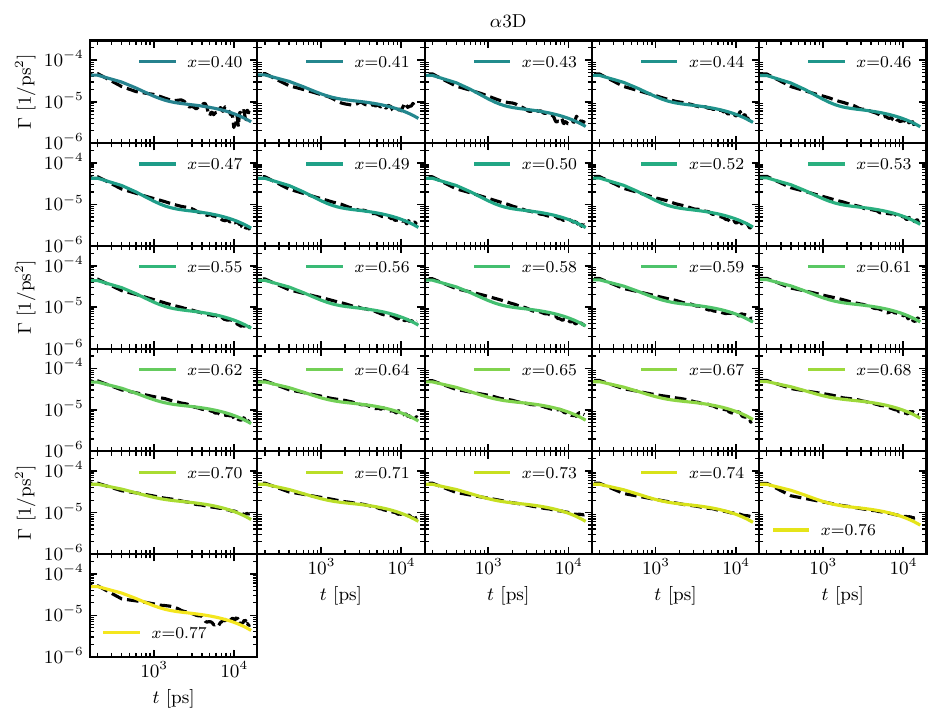} 
    \caption{
 We compare the numerically extracted kernels $\Gamma(t, x)$ (coloured lines) to the
 corresponding fits for Protein G.
 }%
        \label{fig:fits_for_Protein_G}%
    \end{figure}

\begin{figure}
    \centering
    \includegraphics[trim={0 0 0 0},clip]{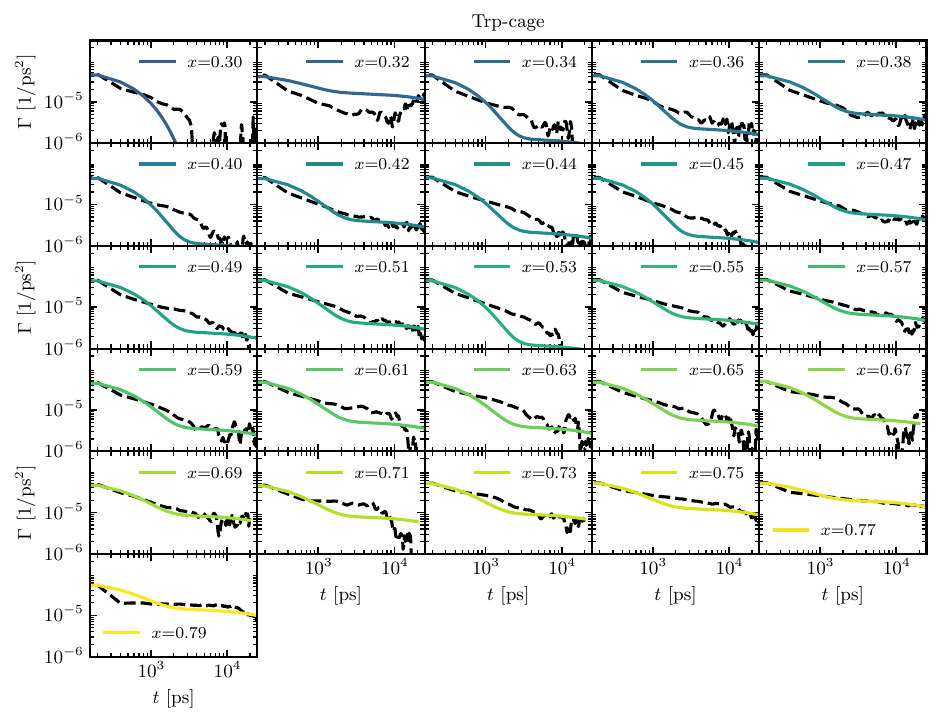} 
    \caption{
 We compare the numerically extracted kernels $\Gamma(t, x)$ (coloured lines) to the
 corresponding fits for the WW domain.
 }%
    \label{fig:fits_for_WWDomain}%
\end{figure}

We fit the numerical kernels to a multi-exponential form given by Eq.~\ref{eq:fit} in the main text.
To help the fit converge, we start from a fit to the kernel in each bin $x_k$ individually, i.e., we obtain
one set of $\tau_i(x), \gamma_i(x)$ for $i \in (1, 2)$ for each bin $x_k$ by using the 'Powell' method implemented by \texttt{scipy.optimize.minimize}~\cite{2020SciPy-NMeth}
to minimize the mean-squared loss
\begin{equation}
    \mathcal{L}(\tau_i(x_k), \gamma_i(x_k)) \propto \sum_{t < \tau_{\mathrm{max}}(x_k)}
 \left( \Gamma(t, x_k) - \sum_{i=1}^{2} \frac{\gamma_i(x_k)}{\tau_i(x_k)} e^{-t / \tau_i(x_k)}\right)^2,
    \label{eq:loss1}
\end{equation}
where the parameter $\tau_{\mathrm{max}}(x)$ is chosen to be 4, 16, 20, 30 or 40 $\: \mathrm{ns}$, as shown in Figure~\ref{fig:shaw_ends},
determined by how many steps the extraction could resolve due to numerical accuracy.
We then take the median of $\tau_i(x)$ over $x$ to obtain a starting value to fit the final form
of the  multi-exponential kernel
\begin{equation}
    \Gamma_{\mathrm{fit}}(t, \tau_i, \gamma_i(x)) = \sum_{i=1}^{2} \frac{\gamma_i(x)}{\tau_i} e^{-t / \tau_i}.
    \label{eq:kernel_embed}
\end{equation}
We again use the 'Powell' method implemented by \texttt{scipy.optimize.minimize}~\cite{2020SciPy-NMeth} to minimize the mean-squared
loss over all times and bins $x_k$
\begin{equation}
    \mathcal{L}(\tau_i, \gamma_i(\cdot)) \propto \sum_{x_k} \sum_{t < \tau_{\mathrm{max}}(x_k)}
 \left( \Gamma(t, x_k) - \Gamma(t, \tau_i, \gamma_i(x_k)) \right)^2.
    \label{eq:loss2}
\end{equation}
We alternate between two fitting modes:
We either change $\gamma_i(x_k)$ for each bin $x_k$ while keeping
$\tau_i$ constant or change $\tau_i$ while keeping $\gamma_i(x_k)$
constant. We repeat the process five times for each mode until convergence is achieved.
Figures~\ref{fig:fits_for_Villin}-\ref{fig:fits_for_WWDomain} show the resulting fits for all
proteins considered in this paper.


\FloatBarrier
\section{\label{si:embedding}Markovian Embedding for RC-Dependent Friction  Kernels}
We here demonstrate the validity of the embedding equations of motion from the main text (Eq.~\ref{eq:eoms})
\begin{equation}
    \begin{split}
        \dot{x}(t) &= v(t), \\
 \dot{v}(t) &= -\frac{1}{m(x(t))}\dv{x(t)} U_{\text{eff}}(x(t), v(t)) + \sum_i \frac{y_i(t)}{\tau_i}, \\
        \dot{y}_i(t) &= -\frac{y_i(t)}{\tau_i} - \bar{\gamma}_i(x(t)) m(x(t)) v(t) + \frac{\bar{\gamma}_i'(x(t))}{2 \bar{\gamma}_i(x(t))} v(t) y_i(t) +\sqrt{k_{\mathrm{B}}T \bar{\gamma}_i(x(t))} \eta_i(t), \\
    \end{split}
    \label{eq:eoms_si}
\end{equation}
with $\langle \eta_i(t) \rangle = 0$, $\langle \eta_i(t) \eta_j(t' )\rangle = 2 \delta_{ij}
\delta(t - t')$ and $\bar{\gamma}_i'(x) = \dv{x} \bar{\gamma}_i(x)$, the spatial derivative of the friction
profile. The term $\frac{\bar{\gamma}_i'(x(t))}{2 \bar{\gamma}_i(x(t))} v(t) y_i(t)$ ensures that the
GLE simulation samples the joint stationary distribution that includes the auxiliary variables $\vec{y}$
\begin{equation}
 P_{\mathrm{aux}}(x, v, \vec{y}) \propto \exp{-\frac{U_{\text{pmf}}(x)}{k_{\mathrm{B}}T} - \frac{m(x)}{2 k_{\mathrm{B}}T} v^2 - \sum_i \left( \frac{1}{2 k_{\mathrm{B}}T \tau_i \bar{\gamma}_i(x)} y_i^2 + \frac{1}{2}\ln(\bar{\gamma}_i(x)) \right) + \frac{1}{2}\ln(m(x))}.
    \label{eq:joint}
\end{equation}
We have constructed Eq.~\ref{eq:joint} such that it has the marginal distributions
\begin{equation}
 \int \dd \vec{y} \: P_{\mathrm{aux}}(x, v, \vec{y}) = P(x, v) \propto \sqrt{m(x)} P_x(x) \exp{-\frac{m(x)}{2 k_{\mathrm{B}}T} v^2},
 \label{eq:marginal_const_v}
\end{equation}
and
\begin{equation}
 \int \dd v \: \dd \vec{y} \: P_{\mathrm{aux}}(x, v, \vec{y}) = P_x(x) \propto \exp{-\frac{U_{\mathrm{pmf}}(x)}{k_{\mathrm{B}}T}}.
 \label{eq:marginal_x}
\end{equation}
When solving Eq.~\ref{eq:eoms_si} for $v(t)$, one obtains a GLE with the form of Eq.~\ref{eq:nonlinGLE} and
the kernel
\begin{equation}
    \Gamma_{\mathrm{sim}}(t - s, x(s)) = m(x(s)) \sum_i \frac{\bar{\gamma}_i(x(s))}{\tau_i} e^{-(t - s) / \tau_i},
    \label{eq:kernel_sim}
\end{equation}
in good approximation.
Consequently, to use the fits in Eq.~\ref{eq:fit} when simulating Eq.~\ref{eq:eoms_si}, we set $\bar{\gamma}_i(x) = \gamma_i(x) / m(x)$.
In Refs.~\cite{ayazGeneralizedLangevinEquation2022,ayazSelfconsistentMarkovianEmbedding2022}, a
relation for the orthogonal force correlation $\langle F_R(0) F_R(t) \rangle$ is given. We do not test if this
relation holds for our embedding.
%
\subsection{\label{si:fokk}Equilibrium Distribution of the RC-Dependent Friction Markovian Embedding}
To derive the Fokker-Planck equation for our system, we follow Ref.~\cite{riskenFokkerPlanckEquationMethods1996}.
We define the vectorial Langevin equation
\begin{equation}
    \dot{\vec{q}}_i(t) = h_i(\vec{q}(t)) + \sum_{j=1}^{N + 2} g_{ij}(\vec{q}(t)) \eta_j(t),
\end{equation}
with $\vec{q}(t) = (x(t), v(t), y_1(t), \dots, y_N(t))$. For the Langevin equation~\ref{eq:eoms_si}, we obtain
\begin{equation}
    \begin{split}
 h(\vec{q}(t)) =
        \begin{bmatrix}
 v(t) \\
 -\frac{1}{m(x(t))}\dv{x(t)} U_{\text{eff}}(x(t), v(t)) + \sum_{i=1}^N \frac{y_i(t)}{\tau_i} \\
 -\frac{y_1(t)}{\tau_1} - \bar{\gamma}_1(x(t)) m(x(t)) v(t) + \frac{\bar{\gamma}_1'(x(t))}{2 \bar{\gamma}_1(x(t))} v(t) y_1(t) \\
        \vdots \\
 -\frac{y_{N}(t)}{\tau_{N}} - \bar{\gamma}_{N}(x(t)) m(x(t)) v(t) + \frac{\bar{\gamma}_{N}'(x(t))}{2 \bar{\gamma}_{N}(x(t))} v(t) y_{N}(t)
        \end{bmatrix},
    \end{split}
\end{equation}
and
\begin{equation}
    \begin{split}
 g(\vec{q}(t)) =
        \begin{bmatrix}
        0 & 0 & 0 & \cdots & 0 \\
        0      & 0 & 0 & \cdots & 0 \\
        0      & 0 & \sqrt{k_{\mathrm{B}}T \bar{\gamma}_1(x(t))} & \cdots & 0 \\
        \vdots & \vdots & \vdots & \ddots & \vdots \\
        0      & 0 & 0 & \cdots & \sqrt{k_{\mathrm{B}}T \bar{\gamma}_N(x(t))} \\
        \end{bmatrix}.
    \end{split}
\end{equation}
Next, we consider the Kramers-Moyal coefficients
\begin{equation}
    \begin{split}
 D_i^{(1)}(\vec{q}(t)) &= h_i(\vec{q}(t)) + \sum_{j=1}^{N + 2} \sum_{k=1}^{N + 2} g_{kj}(\vec{q}(t)) \pderivative{q_k} g_{ij}(\vec{q}(t)), \\
 D_{ij}^{(2)}(\vec{q}(t)) &= \sum_{k=1}^{N + 2} g_{ik}(\vec{q}(t)) g_{jk}(\vec{q}(t)),
    \end{split}
\end{equation}
where $D_i^{(1)}(\vec{q}(t))$ are the first-order and $D_{ij}^{(2)}(\vec{q}(t))$ the second-order
Kramers-Moyal coefficients (see Ref.~\cite{riskenFokkerPlanckEquationMethods1996}, pp. 49-50, Eqs. 3.93-3.94).
For the equations of motion in Eq.~\ref{eq:eoms_si}, the non-zero Kramers-Moyal coefficients
are
\begin{equation}
    \begin{split}
 D_x^{(1)} &= v(t), \\
 D_v^{(1)} &= -\frac{1}{m(x)} \left[ \dv{x(t)}U_{\text{eff}}(x(t), v(t)) \right] + \sum_i \frac{y_i(t)}{\tau_i}, \\
 &= -\frac{1}{m(x)} \left( U'_{\text{pmf}}(x) + \frac{m'(x)}{2} v^2 + \frac{k_{\mathrm{B}}T m'(x)}{2 m(x)} \right) + \sum_i \frac{y_i(t)}{\tau_i}, \\
 D_{y_i}^{(1)} &= -\frac{y_i(t)}{\tau_i} - \bar{\gamma}_i(x(t)) m(x(t)) v(t) + \frac{\bar{\gamma}_i'(x(t))}{2 \bar{\gamma}_i(x(t))} v(t) y_i(t), \\
 D_{y_i y_i}^{(2)} &= k_{\mathrm{B}}T \bar{\gamma}_i(x(t)).
    \end{split}
    \label{eq:KMcoeffs}
\end{equation}
We obtain the Fokker-Planck equation from the first and second order Kramers-Moyal coefficients,
yielding~\cite{riskenFokkerPlanckEquationMethods1996}

\begin{equation}
    \begin{split}
        \pderivative{t} P(\vec{q},t)
        =
        &-\sum_{i=1}^{N+2}\pderivative{q_i}
        \left[D_i^{(1)}(\vec{q}) P(\vec{q},t)\right]
        + \sum_{i=1}^{N+2}\sum_{j=1}^{N+2}
        \pdv{q_i}\pdv{q_j}
        \left[D_{ij}^{(2)}(\vec{q}) P(\vec{q},t)\right] \\
        =
        &-\pderivative{x}\left[D_x^{(1)} P(\vec{q},t)\right]
        -\pderivative{v}\left[D_v^{(1)} P(\vec{q},t)\right] \\
        &+ \sum_{i=1}^{N}
        \left(
        -\pderivative{y_i}\left[D_{y_i}^{(1)} P(\vec{q},t)\right]
        + \pderivative[2]{y_i}\left[D_{y_i y_i}^{(2)} P(\vec{q},t)\right]
        \right).
    \end{split}
    \label{eq:fp}
\end{equation}
Inserting the Kramers-Moyal coefficients from Eq.~\ref{eq:KMcoeffs} into Eq.~\ref{eq:fp} and using
the distribution $P_{\mathrm{aux}}(\vec{q})$ in Eq.~\ref{eq:joint} then leads to
\begin{equation}
    \begin{split}
 -\pderivative{x} \left( D_x^{(1)} P_{\mathrm{aux}}(\vec{q})\right) &= -v \left(-\frac{U'_{\text{pmf}}(x)}{k_{\mathrm{B}}T} - \frac{m'(x)}{2 k_{\mathrm{B}}T} v^2 + \frac{m'(x)}{2 m(x)} + \sum_i \left(\frac{\bar{\gamma}_i'(x)}{2 k_{\mathrm{B}}T \tau_i \bar{\gamma}_i(x)^2} y_i^2 - \frac{\bar{\gamma}_i'(x)}{2 \bar{\gamma}_i(x)} \right) \right) P_{\mathrm{aux}}(\vec{q}), \\
        &= \left( \frac{U'_{\text{pmf}}(x)}{k_{\mathrm{B}}T} v + \frac{m'(x)}{2 k_{\mathrm{B}}T} v^3 - \frac{m'(x)}{2 m(x)} v + \sum_i \left(-\frac{\bar{\gamma}_i'(x)}{2 k_{\mathrm{B}}T \tau_i \bar{\gamma}_i(x)^2} v y_i^2 + \frac{\bar{\gamma}_i'(x)}{2 \bar{\gamma}_i(x)} v\right) \right) P_{\mathrm{aux}}(\vec{q}), \\
 -\pderivative{v} \left( D_v^{(1)} P_{\mathrm{aux}}(\vec{q})\right) &= -\left(-\frac{m(x)}{k_{\mathrm{B}}T} v  \left( -\frac{U'_{\text{pmf}}(x)}{m(x)} - \frac{m'(x)}{2 m(x)} v^2 - \frac{k_{\mathrm{B}}T m'(x)}{2 m(x)^2} + \sum_i \frac{y_i}{\tau_i} \right) - \frac{m'(x)}{m(x)}v \right) P_{\mathrm{aux}}(\vec{q}), \\
        &= \left(-\frac{U'_{\text{pmf}}(x)}{k_{\mathrm{B}}T} v - \frac{m'(x)}{2 k_{\mathrm{B}}T} v^3 + \frac{m'(x)}{2 m(x)} v +\sum_i \frac{m(x)}{k_{\mathrm{B}}T \tau_i} vy_i\right) P_{\mathrm{aux}}(\vec{q}), \\
 -\pderivative{y_i} \left( D_{y_i}^{(1)} P_{\mathrm{aux}}(\vec{q})\right) &= \left(\frac{1}{\tau_i} -\frac{1}{k_{\mathrm{B}}T \tau_i^2 \bar{\gamma}_i(x)} y_i^2 -\frac{m(x)}{k_{\mathrm{B}}T \tau_i} vy_i -\frac{\bar{\gamma}_i'(x)}{2 \bar{\gamma}_i(x)} v + \frac{\bar{\gamma}_i'(x)}{2 k_{\mathrm{B}}T \tau_i \bar{\gamma}_i(x)^2} v y_i^2 \right)  P_{\mathrm{aux}}(\vec{q}), \\
        \pderivative[2]{y_i} \left( D_{y_i y_i}^{(2)} P_{\mathrm{aux}}(\vec{q})\right) &= k_{\mathrm{B}}T \bar{\gamma}_i(x) \left( \frac{1}{\left(k_{\mathrm{B}}T \tau_i \bar{\gamma}_i(x)\right)^2} y_i^2 - \frac{1}{k_{\mathrm{B}}T \tau_i \bar{\gamma}_i(x)} \right) P_{\mathrm{aux}}(\vec{q}), \\
        &= \left( \frac{1}{k_{\mathrm{B}}T \tau_i^2 \bar{\gamma}_i(x)} y_i^2 - \frac{1}{\tau_i} \right) P_{\mathrm{aux}}(\vec{q}). \\
    \end{split}
    \label{eq:fp_calc}
\end{equation}
Inserting all terms in Eq.~\ref{eq:fp_calc} into Eq.~\ref{eq:fp}, we obtain
\begin{equation}
    \pderivative{t} P_{\mathrm{aux}}(\vec{q}, t) = 0,
\end{equation}
for the distribution $P_{\mathrm{aux}}(\vec{q})$ given in Eq.~\ref{eq:joint}, demonstrating that the equations of motion in Eq.~\ref{eq:eoms_si} sample the stationary distribution in Eq.~\ref{eq:joint}.
\subsection{Validating the RC-Dependent Friction Markovian Embedding}
\begin{figure}
    \centering
    \includegraphics{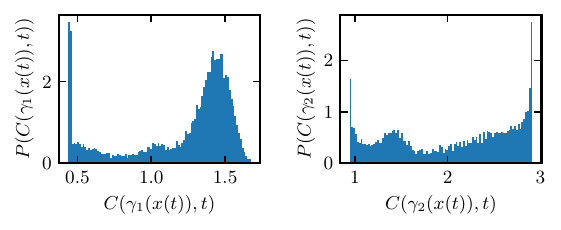}
    \caption{We evaluate $C(\bar{\gamma}_i(x(t)), t)$ defined in Eq.~\ref{eq:cfacs}
 for a 250 $\mu$s GLE simulation of Villin using the equations of motion in Eq.~\ref{eq:eoms_si}.
 We show a histogram of $C(\bar{\gamma}_i(x(t)), t)$ for 1000 different $t$ equally spaced along the trajectory
 for both friction components $\bar{\gamma}_1(x)$ and $\bar{\gamma}_2(x)$.
 }%
    \label{fig:4}
\end{figure}

Here we show that the embedding containing $\bar{\gamma}_i(x)$ in Eq.~\ref{eq:eoms_si} corresponds to the form of the kernel in the
GLE in Eq.~\ref{eq:nonlinGLE} in the main text to a good approximation.
The solution of the first-order inhomogeneous differential equation
\begin{equation}
    \dot{y}_i(t) + p_i(t) y_i(t) = f_i(t),
\end{equation}
is given by
\begin{equation}
\begin{split}
 y_i(t) &= y_i(t_0) e^{-\int_{t_0}^t \dd s \: p_i(s)} + e^{-\int_{t_0}^t \dd s' \: p_i(s')} \int_{t_0}^t \dd s f_i(s) e^{\int_{t_0}^s \dd s' \: p_i(s')}, \\
    &= y_i(t_0) e^{-\int_{t_0}^t \dd s \: p_i(s)} + \int_{t_0}^t \dd s f_i(s) e^{\int_{t_0}^s \dd s' \: p_i(s') - \int_{t_0}^t \dd s' \: p_i(s')}, \\
    &= y_i(t_0) e^{-\int_{t_0}^t \dd s \: p_i(s)} + \int_{t_0}^t \dd s f_i(s) e^{-\int_{s}^t \dd s' \: p_i(s')}.
\end{split}
\label{eq:solyfirst}
\end{equation}
From the equations of motion in Eq.~\ref{eq:eoms_si}, we obtain
\begin{equation}
    \begin{split}
 p_i(t) &= \frac{1}{\tau_i} - \frac{\bar{\gamma}_i'(x(t))}{2 \bar{\gamma}_i(x(t))} v(t), \\
 f_i(t) &= -\bar{\gamma}_i(x(t)) m(x(t)) v(t)  + \sqrt{ k_{\mathrm{B}}T \bar{\gamma}_i(x(t))} \eta_i(t).
    \end{split}
    \label{eq:ps_fs_of_y}
\end{equation}
From this, we obtain
\begin{equation}
 e^{-\int_{t_0}^t \dd s \: p_i(s)} = e^{-(t - t_0) / \tau_i} e^{\int_{t_0}^t \dd s \: \frac{\bar{\gamma}_i'(x(s))}{2 \bar{\gamma}_i(x(s))} v(s)}.
\end{equation}
Inserting this into Eq.~\ref{eq:solyfirst} and sending $t_0$ to $-\infty$, we obtain
\begin{equation}
 y_i(t) = \int_{-\infty}^t \dd s e^{-(t - s) / \tau_i} e^{\int_{s}^t \dd s' \: \frac{\bar{\gamma}_i'(x(s'))}{2 \bar{\gamma}_i(x(s'))} v(s')} f_i(s).
    \label{eq:soly2}
\end{equation}
Consequently, the kernel we obtain is
\begin{equation}
    \begin{split}
        \Gamma_{\mathrm{full}}(t - s, x(s)) &= m(x(s)) \sum_i e^{\int_{s}^t \dd s' \: \frac{\bar{\gamma}_i'(x(s'))}{2\bar{\gamma}_i(x(s'))} v(s')} \frac{\bar{\gamma}_i(x(s))}{\tau_i} e^{-(t - s) / \tau_i}, \\
         &= m(x(s)) \sum_i e^{\int_{s}^t \dd s' \: \frac{\bar{\gamma}_i'(x(s'))}{2 \bar{\gamma}_i(x(s'))} v(s')} \bar{\Gamma}_{\mathrm{sim}}(t - s, x(s)).
    \end{split}
    \label{eq:kernel_sim_full}
\end{equation}
In order for $\Gamma_{\mathrm{full}}(t - s, x(s))$ to approximately correspond to
$\Gamma_{\mathrm{sim}}(t - s, x(s))$ in Eq.~\ref{eq:kernel_sim}, we need to show that $e^{\int_{s}^t
\dd s' \: \frac{\bar{\gamma}_i'(x(s'))}{2 \bar{\gamma}_i(x(s'))} v(s')}$ is close to one. For systems
with small inertial time $m / \gamma$ (overdamped dynamics), $v(t)$ will change much faster than $x(t)$.
Assuming that $v(t)$ equilibrates over an infinitesimal time $\dd s'$ and $v(t)$ and $x(t)$ are
uncorrelated, we can replace $v(t)$ with its mean value $\langle v \rangle$. We then obtain
\begin{equation}
    \begin{split}
    e^{\int_{s}^t \dd s' \: \frac{\bar{\gamma}_i'(x(s'))}{2 \bar{\gamma}_i(x(s'))} v(s')}
    &\approx e^{\langle v \rangle \int_{s}^t \dd s' \: \frac{\bar{\gamma}_i'(x(s'))}{2 \bar{\gamma}_i(x(s'))}} = 1,
    \end{split}
\end{equation}
since $\langle v \rangle = 0$.
To test the approximation, we evaluate
\begin{equation}
    C(\bar{\gamma}_i(x(t)), t) = e^{\int_{-\infty}^t \dd s \: \frac{\bar{\gamma}_i'(x(s))}{2 \bar{\gamma}_i(x(s))} v(s)}
 \label{eq:cfacs}
\end{equation}
numerically for the GLE simulation of the Villin protein simulation shown in Figure~\ref{fig:4}. We
see that the values for $C(\bar{\gamma}_i(x(t)), t)$ are between 0.5 and 3. Thus,
$C(\bar{\gamma}_i(x(t)), t)$ is of the order of one, suggesting that the approximation broadly
holds for the two components of the fit $\bar{\gamma}_1(x)$ and $\bar{\gamma}_2(x)$. However, we come up with a more stringent
estimation for the error in the following. \\

\begin{figure}
    \centering
    \includegraphics{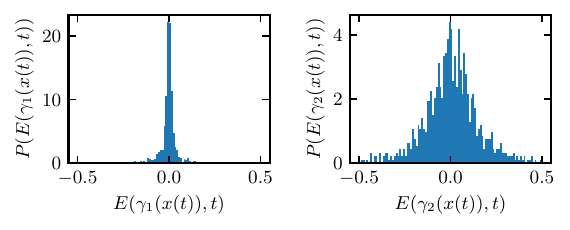}
    \caption{Estimate of the effect the correction factor $\frac{\bar{\gamma}_i'(x(t))}{2 \bar{\gamma}_i(x(t))} v(t) y_i(t)$
 has on the GLE dynamics using $E(\bar{\gamma}_i(t), t)$, defined in Eq.~\ref{eq:relerr},
 for a 250 $\mu s$ GLE simulation of Villin following the equation of motion in Eq.~\ref{eq:eoms_si}.
 We show histograms of $E(\bar{\gamma}_i(t), t)$ for 1000 different $t$ equally spaced along the trajectory
 for both friction components $\bar{\gamma}_1(x)$ and $\bar{\gamma}_2(x)$.
 }%
    \label{fig:5}
\end{figure}

To estimate the error of the GLE simulation, we first consider the
solution for $y_i(t)$ of the embedding system by inserting Eq.~\ref{eq:ps_fs_of_y} into Eq.~\ref{eq:soly2}, yielding
\begin{equation}
 y_i^{\text{embedding}}(t) = \int_{-\infty}^t \dd s e^{-(t - s) / \tau_i} e^{\int_{s}^t \dd s' \: \frac{\bar{\gamma}_i'(x(s'))}{2\bar{\gamma}_i(x(s'))} v(s')} \left( -\bar{\gamma}_i(x(s)) m(x(s)) v(s)  + \sqrt{ k_{\mathrm{B}}T \bar{\gamma}_i(x(s))} \eta_i(s) \right).
    \label{eq:ycorr}
\end{equation}
Next, we set the terms $\frac{\bar{\gamma}_i'(x(s'))}{2\bar{\gamma}_i(x(s'))} v(s')$ to zero and obtain
\begin{equation}
 y_i^{\text{reference}}(t) = \int_{-\infty}^t \dd s e^{-(t - s) / \tau_i} \left( -\bar{\gamma}_i(x(s)) m(x(s)) v(s)  + \sqrt{ k_{\mathrm{B}}T \bar{\gamma}_i(x(s))} \eta_i(s) \right),
    \label{eq:ybase}
\end{equation}
enabling us to compute the effect $\frac{\bar{\gamma}_i'(x(s'))}{2\bar{\gamma}_i(x(s'))} v(s')$
has on the kernel during the embedding simulation.
The result of this computation, compared to $C(\bar{\gamma}_i(x(t)), t)$, provides an estimation of the error defined as
\begin{equation}
 E(\bar{\gamma}_i(x(t)), t) = \frac{y_i^{\text{embedding}}(t)}{y_i^{\text{reference}}(t)} - 1,
    \label{eq:relerr}
\end{equation}
which is defined such that it goes to zero when the kernels in Eq.~\ref{eq:kernel_sim} and Eq.~\ref{eq:kernel_sim_full} are exactly
the same.
Figure~\ref{fig:5} shows that the error $E(\bar{\gamma}_i(x(t)), t)$ is small for most frames of the GLE simulation, demonstrating
that the embedding can be used for position-dependent GLE simulations and approximately corresponds to the kernel
in Eq.~\ref{eq:kernel_sim}.

\FloatBarrier
%
\subsection{\label{si:sims}GLE Simulations}
For all GLE simulations, we use the fourth-order Runge-Kutta integrator. GLE simulations without position-dependent
friction and mass (Eq.~\ref{eq:GLE}) are done with the equations of motion
\begin{equation}
    \begin{split}
        \dot{x}(t) &= v(t), \\
 m \dot{v}(t) &= -U_{\text{pmf}}'(x(t)) + \sum_{i=1}^{N} \frac{\gamma_i}{\tau_i} \left(y_i(t) - x(t)\right), \\
        \dot{y}_i(t) &= -\frac{1}{\tau_i} \left(y_i(t) - x(t)\right) + \sqrt{\frac{k_{\text{B}}T}{\gamma_i}} \eta_i(t), \\
    \end{split}
\end{equation}
which represents a kernel of the form of Eq.~\ref{eq:kernel_ben} in the main
text~\cite{ayazNonMarkovianModelingProtein2021}. For all GLE simulations, we use a time step of
$0.5\:\text{ps}$ and $5\times 10^{10}$ steps. The RC-independent masses are listed in
Table~\ref{tab:protein_masses}, and computed using $m = k_{\mathrm{B}}T / \langle v^2 \rangle$,
where the velocities are calculated using the numerical gradient in Eq.~\ref{eq:gradient_halfstep}.

\begin{table}[htbp]
\centering
\caption{Masses used in RC-independent GLE simulations.}
\begin{tabular}{|l|r|}
\hline
Protein & Mass [u] \\
\hline
Trp-cage & $1.094 \cdot 10^{8}$ \\
Villin & $2.136 \cdot 10^{8}$ \\
WW domain & $2.885 \cdot 10^{8}$ \\
Protein G & $7.648 \cdot 10^{8}$ \\
$\alpha_3D$ & $1.123 \cdot 10^{9}$ \\
$\lambda$-repressor & $1.173 \cdot 10^{9}$ \\
\hline
\end{tabular}
\label{tab:protein_masses}
\end{table}

\begin{figure}
    \centering
    \includegraphics{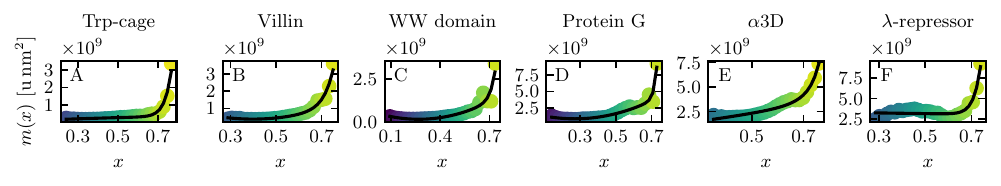} 
    \caption{We compare the fit for $m(x)$ according to Eq.~\ref{eq:si_fit_mass} used in the GLE simulations to the numerical data
 computed from the MD simulations via Eq.~\ref{eq:mass_calc}. Table~\ref{table:cond_mass_fit} lists the corresponding parameters.
 }%
    \label{figsi:cond_mass_fit}%
\end{figure}

\begin{figure}
    \centering
    \includegraphics{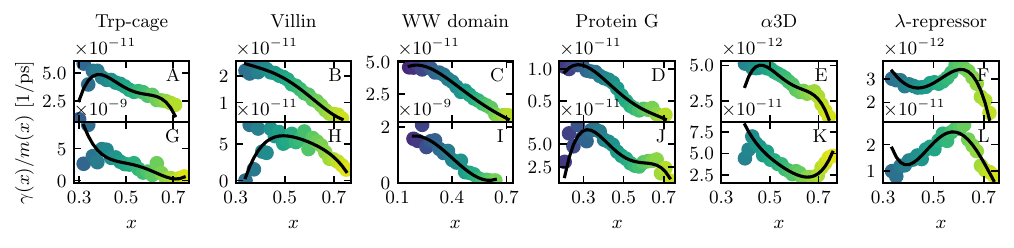} 
    \caption{
 We compare the fits for $\gamma_i(x) / m(x)$ according to Eqs.~\ref{eq:si_fit_gam} and \ref{eq:si_fit_mass}
 used in the GLE simulations to the numerical data
 extracted from the MD simulations. The upper row \textbf{A-F} shows the fits for $\gamma_1(x)$
 and the lower row \textbf{G-L} shows the fits for $\gamma_2(x)$. Table~\ref{table:si_fit_gam} lists
 the corresponding parameters.
 }%
    \label{fig:si_fit_gam}%
\end{figure}

%
For simulations with a position-dependent mass, we represent $m(x)$ in Eq.~\ref{eq:eoms_si} by the following fit function
\begin{equation}
 m_{\mathrm{fit}}(x, \theta_j) = \theta_1 \exp\left( \left(\frac{x}{\theta_2}\right)^{\theta_3} \right) + \theta_4 + \theta_5 x + \theta_6 x^2,
    \label{eq:si_fit_mass}
\end{equation}
with the corresponding derivative
\begin{equation}
    m_{\mathrm{fit}}'(x,\theta_j)
    = \frac{\theta_1 \theta_3}{\theta_2} \left(\frac{x}{\theta_2}\right)^{\theta_3-1}
    \exp\!\left(\left(\frac{x}{\theta_2}\right)^{\theta_3}\right)
    + \theta_5 + 2\theta_6 x.
\end{equation}
where we use \texttt{scipy.optimize.curve\_fit}~\cite{2020SciPy-NMeth}~to obtain the parameters $\theta_j$ from the numerical
mass profiles using Eq.~\ref{eq:si_fit_mass}. Figure~\ref{figsi:cond_mass_fit} compares the fits to
the numerical data. Table~\ref{table:cond_mass_fit} lists the fitting parameters $\theta_j$.
\\
For GLE simulations with position-dependent friction, we represent the position-dependent friction profiles
$\bar{\gamma}_i(x)$ in Eq.~\ref{eq:eoms_si} by the fit function
\begin{equation}
    \bar{\gamma}_{\mathrm{fit}}(x, \theta_{i,j}, x^{\mathrm{min}}_i, x^{\mathrm{max}}_i)_i = \sum_{j = 1}^5 \theta_{i,j} c(x, x^{\mathrm{min}}_i, x^{\mathrm{max}}_i)^{j - 1},
    \label{eq:si_fit_gam}
\end{equation}
where $c(x, x^{\mathrm{min}}_i, x^{\mathrm{max}}_i) = \max(\min(x, x^{\mathrm{max}}), x^{\mathrm{min}})$ restricts the position $x$ to the
interval $(x^{\mathrm{min}}, x^{\mathrm{max}})$.
The gradient follows as
\begin{equation}
    \bar{\gamma}_{\mathrm{fit}}'(x,\theta_{i,j},x_i^{\mathrm{min}},x_i^{\mathrm{max}})
    =
    \begin{cases}
        \displaystyle
        \sum_{j=2}^{5} (j-1)\theta_{i,j}\,x^{j-2},
        & x_i^{\mathrm{min}} < x < x_i^{\mathrm{max}}, \\
        0,
        & \text{otherwise}.
    \end{cases}
    \label{eq:si_fit_gam_der}
\end{equation}
We use \texttt{scipy.optimize.curve\_fit}~\cite{2020SciPy-NMeth}~to obtain the parameters $\theta_{i,j}$ from the numerical
friction profiles using Eq.~\ref{eq:si_fit_gam}. Figure~\ref{fig:si_fit_gam} compares the fits to
the numerical data. Table~\ref{table:si_fit_gam} lists the fitting parameters $\theta_i$, as well as
$x^{\mathrm{min}}_i$ and $x^{\mathrm{max}}_i$.

Whenever we estimate mean first-passage times $\tau_{\mathrm{MFP}}(x_s, x_f)$ from GLE or MD
simulation for a starting point $x_s$ and a final point $x_f$, we identify each time the
trajectory crosses $x_s$ before first reaching $x_f$, and then identify the transition time of the
transition events as the time differences between the instances of $x_s$ prior to the time of
arriving at the final state $x_f$. We repeat this procedure for each occurrence of $x_f$ in the
trajectory. We calculate $\tau_{\mathrm{MFP}}(x_s, x_f)$ by averaging over all identified transition
times.

\begin{table}
    \centering
    \caption{We show the parameters $\theta_j$ for the fit of $m(x)$ in Eq.~\ref{eq:si_fit_mass}.}
    \begin{tabular}{| c | c | c | c | c | c | c |}
        \hline
        & $\theta_1$ [u $\:$nm$^2$] & $\theta_2$ & $\theta_3$ & $\theta_4$ [u $\:$nm$^2$] & $\theta_5$ [u $\:$nm$^2$] & $\theta_6$ [u $\:$nm$^2$] \\ \hline
 Villin & 1.724e-96 & 5.451e-10 & 2.607e-01 & 3.671e+08 & -1.856e+09 & 2.868e+09 \\
        $\lambda$-repressor & 1.206e-01 & 2.267e-01 & 2.547e+00 & 4.233e+08 & -1.608e+09 & 1.874e+09 \\
        $\alpha$3D & 6.059e-155 & 3.284e-18 & 1.484e-01 & 9.783e+08 & -3.952e+09 & 6.081e+09 \\
 WW domain & 4.014e-217 & 7.218e-28 & 1.005e-01 & 1.113e+09 & -4.330e+09 & 4.813e+09 \\
 Trp-cage & 1.406e-110 & 4.827e-13 & 1.998e-01 & 9.163e+08 & -2.497e+09 & 4.344e+09 \\
 Protein G & 1.119e+09 & 7.400e-01 & 1.868e+01 & 1.000e+00 & 1.000e+00 & -4.480e+08 \\
        \hline
    \end{tabular}
    \label{table:cond_mass_fit}
\end{table}

\begin{table}
    \centering
    \caption{We show the parameters $\theta_{i,j}$ for the fit of $\gamma_i(x)$ in Eq.~\ref{eq:si_fit_gam}
 as well as the corresponding memory time scales $\tau_i$ and the range in which the fit is evaluated
    $(x^{\mathrm{min}}_i, x^{\mathrm{max}}_i)$.}
    \begin{tabular}{| c | c | c | c | c | c | c | c | c |}
        \hline
        & $\theta_{1,1}$ [1/ps] & $\theta_{1,2}$ [1/ps] & $\theta_{1,3}$ [1/ps] & $\theta_{1,4}$ [1/ps] & $\theta_{1,5}$ [1/ps] & $x^{\mathrm{min}}_1$ & $x^{\mathrm{max}}_1$ & $\tau_1$ [ps] \\ \hline
 Villin & -3.374e-10 & 7.699e-09 & -2.760e-08 & 3.649e-08 & -1.660e-08 & 0.16 & 0.73 & 3.847e+02 \\
        $\lambda$-repressor & -6.607e-09 & 5.694e-08 & -1.679e-07 & 2.136e-07 & -1.006e-07 & 0.30 & 0.73 & 5.073e+02 \\
        $\alpha$3D & -1.874e-10 & 2.338e-09 & -7.408e-09 & 9.368e-09 & -4.241e-09 & 0.21 & 0.73 & 3.556e+02 \\
 WW domain & -2.358e-09 & 1.594e-08 & -3.547e-08 & 3.225e-08 & -1.017e-08 & 0.34 & 0.75 & 3.435e+02 \\
 Trp-cage & -1.452e-09 & 1.039e-08 & -2.666e-08 & 2.978e-08 & -1.233e-08 & 0.40 & 0.76 & 3.754e+02 \\
 Protein G & -1.803e-10 & 2.029e-09 & -7.350e-09 & 1.130e-08 & -6.244e-09 & 0.34 & 0.74 & 3.894e+02 \\
        \hline
        & $\theta_{2,1}$ [1/ps] & $\theta_{2,2}$ [1/ps] & $\theta_{2,3}$ [1/ps] & $\theta_{2,4}$ [1/ps] & $\theta_{2,5}$ [1/ps] & $x^{\mathrm{min}}_2$ & $x^{\mathrm{max}}_2$ & $\tau_2$ [ps] \\ \hline
 Villin & 8.452e-09 & -7.758e-08 & 2.970e-07 & -4.511e-07 & 2.324e-07 & 0.16 & 0.76 & 4.670e+04 \\
        $\lambda$-repressor & -1.169e-07 & 9.145e-07 & -2.569e-06 & 3.142e-06 & -1.414e-06 & 0.32 & 0.77 & 6.428e+04 \\
        $\alpha$3D & -5.670e-09 & 5.640e-08 & -1.831e-07 & 2.507e-07 & -1.243e-07 & 0.21 & 0.75 & 1.473e+04 \\
 WW domain & 8.157e-09 & -5.204e-08 & 1.287e-07 & -1.337e-07 & 4.766e-08 & 0.34 & 0.77 & 1.241e+04 \\
 Trp-cage & -2.036e-08 & 1.522e-07 & -4.180e-07 & 5.035e-07 & -2.245e-07 & 0.40 & 0.77 & 1.421e+04 \\
 Protein G & 9.237e-09 & -6.982e-08 & 1.928e-07 & -2.261e-07 & 9.519e-08 & 0.34 & 0.76 & 1.243e+04 \\
        \hline
    \end{tabular}
    \label{table:si_fit_gam}
\end{table}

\FloatBarrier
\subsection{Alternate GLEs with Different Effective Potential}
We note that the embedding in Eq.~\ref{eq:eoms_si} for the GLE in Eq.~\ref{eq:nonlinGLE} is fully defined by the choice of
distributions for $x$ and $v$ in Eqs.~\ref{eq:marginal_x} and~\ref{eq:marginal_const_v} and the functional form of the kernel in Eq.~\ref{eq:kernel_sim}.
It is possible to derive non-linear GLEs with a different effective potential. One such GLE takes
the same form as Eq.~\ref{eq:nonlinGLE}, but has the slightly modified effective potential~\cite{vroylandtDerivationGeneralizedLangevin2022,heryGeneralizedLangevinEquations}
\begin{equation}
 U_{\text{eff}*}(x) = U_{\mathrm{pmf}}(x) + k_{\mathrm{B}}T \ln m(x).
    \label{eq:gle_ueff1}
\end{equation}
As an ansatz to derive an embedding, we start with the equations of motion in Eq.~\ref{eq:eoms_si}
\begin{equation}
    \begin{split}
        \dot{x}(t) &= v(t), \\
 \dot{v}(t) &= -\frac{1}{m(x(t))}\dv{x(t)} U_{\text{eff}*}(x(t)) + \sum_i \frac{y_i(t)}{\tau_i} \textcolor{blue}{+ c(x(t), v(t))}, \\
        \dot{y}_i(t) &= -\frac{y_i(t)}{\tau_i} - \bar{\gamma}_i(x(t)) m(x(t)) v(t) + \frac{\bar{\gamma}_i'(x(t))}{2 \bar{\gamma}_i(x(t))} v(t) y_i(t) +\sqrt{k_{\mathrm{B}}T \bar{\gamma}_i(x(t))} \eta_i(t), \\
    \end{split}
    \label{eq:eoms_si_ueff1}
\end{equation}
where we use the effective potential $U_{\text{eff}*}(x)$ and add the correction term $\textcolor{blue}{c(x, v)}$, to be determined below.
To determine $\textcolor{blue}{c(x, v)}$, similar to our previous procedure, we choose the marginal distributions
\begin{equation}
    \int \dd \vec{y} \: P_{\mathrm{aux}}(x, v, \vec{y}) = P(x, v) \propto \sqrt{m(x)} P_x(x) \exp{-\frac{m(x)}{2 k_{\mathrm{B}}T} v^2},
 \label{eq:marginal_const_v_1}
\end{equation}
and
\begin{equation}
 \int \dd v \: \dd \vec{y} \: P_{\mathrm{aux}}(x, v, \vec{y}) = P_{x}(x) \propto \exp{-\frac{U_{\mathrm{pmf}}(x)}{k_{\mathrm{B}}T}},
 \label{eq:marginal_x_1}
\end{equation}
which follows from the full distribution
\begin{equation}
 P_{\mathrm{aux}}(x, v, \vec{y}) \propto \exp{-\frac{U_{\text{pmf}}(x)}{k_{\mathrm{B}}T} - \frac{m(x)}{2 k_{\mathrm{B}}T} v^2 - \sum_i \left( \frac{1}{2 k_{\mathrm{B}}T \tau_i \bar{\gamma}_i(x)} y_i^2 + \frac{1}{2}\ln(\bar{\gamma}_i(x)) \right) + \frac{1}{2}\ln(m(x))}.
    \label{eq:joint_1}
\end{equation}
To ensure that Eq.~\ref{eq:eoms_si_ueff1}
samples this joint distribution in Eq.~\ref{eq:joint_1}, we repeat the same Fokker-Planck analysis as in section~\ref{si:fokk} with $U_{\text{eff}*}(x)$
instead of $U_{\text{eff}}(x, v)$, which yields

\begin{equation}
    \begin{split}
 -\pderivative{x} \left( D_x^{(1)} P_{\mathrm{aux}}(\vec{q})\right) &= -v \left(-\frac{U'_{\text{pmf}}(x)}{k_{\mathrm{B}}T} - \frac{m'(x)}{2 k_{\mathrm{B}}T} v^2 + \frac{m'(x)}{2 m(x)} + \sum_i \left(\frac{\bar{\gamma}_i'(x)}{2 \bar{\gamma}_i(x)} v - \frac{\bar{\gamma}_i'(x)}{2 k_{\mathrm{B}}T \bar{\gamma}_i(x)^2} v y_i^2 \right) \right) P_{\mathrm{aux}}(\vec{q}), \\
        &= \left( \frac{U'_{\text{pmf}}(x)}{k_{\mathrm{B}}T} v + \frac{m'(x)}{2 k_{\mathrm{B}}T} v^3 - \frac{m'(x)}{2 m(x)} v \right) P_{\mathrm{aux}}(\vec{q}), \\
 -\pderivative{v} \left( D_v^{(1)} P_{\mathrm{aux}}(\vec{q})\right) &= -\left(-\frac{m(x)}{k_{\mathrm{B}}T} v  \left( -\frac{U'_{\text{pmf}}(x)}{m(x)} - \frac{k_{\mathrm{B}}T m'(x)}{m(x)^2} + \sum_i \frac{y_i}{\tau_i} \right) \right) P_{\mathrm{aux}}(\vec{q}) - \textcolor{blue}{\pderivative{v} c(x, v)  P_{\mathrm{aux}}(\vec{q})}, \\
        &= \left(-\frac{U'_{\text{pmf}}(x)}{k_{\mathrm{B}}T} v - \frac{m'(x)}{m(x)} v +\sum_i \frac{m(x)}{k_{\mathrm{B}}T \tau_i} vy_i\right) P_{\mathrm{aux}}(\vec{q}) - \textcolor{blue}{\pderivative{v} c(x, v)  P_{\mathrm{aux}}(\vec{q})}, \\
 -\pderivative{y} \left( D_{y_i}^{(1)} P_{\mathrm{aux}}(\vec{q})\right) &= \left(\frac{1}{\tau_i} -\frac{1}{k_{\mathrm{B}}T \tau_i^2 \bar{\gamma}_i(x)} y_i^2 -\frac{m(x)}{k_{\mathrm{B}}T \tau_i} vy_i -\frac{\bar{\gamma}_i'(x)}{2 \bar{\gamma}_i(x)} v + \frac{\bar{\gamma}_i'(x)}{2 k_{\mathrm{B}}T \bar{\gamma}_i(x)^2} v y_i^2 \right)  P_{\mathrm{aux}}(\vec{q}), \\
        \pderivative[2]{y_i} \left( D_{y_iy_i}^{(2)} P_{\mathrm{aux}}(\vec{q})\right) &= k_{\mathrm{B}}T \bar{\gamma}_i(x) \left( \frac{1}{\left(k_{\mathrm{B}}T \tau_i \bar{\gamma}_i(x)\right)^2} y_i^2 - \frac{1}{k_{\mathrm{B}}T \tau_i \bar{\gamma}_i(x)} \right) P_{\mathrm{aux}}(\vec{q}), \\
        &= \left( \frac{1}{k_{\mathrm{B}}T \tau_i^2 \bar{\gamma}_i(x)} y_i^2 - \frac{1}{\tau_i} \right) P_{\mathrm{aux}}(\vec{q}) \\
    \end{split}
    \label{eq:fp_calc_ueff1}
\end{equation}
For the terms in Eq.~\ref{eq:fp_calc_ueff1} to sum up to zero, we find that
\begin{equation}
    \pderivative{v} \left( c(x, v) P_{\mathrm{aux}}(\vec{q}) \right) = \left( \frac{m'(x)}{2 k_{\mathrm{B}}T} v^3 - \frac{3 m'(x)}{2 m(x)}v \right) P_{\mathrm{aux}}(\vec{q}),
    \label{eq:si444}
\end{equation}
has to hold. From that condition we obtain
\begin{equation}
 c(x, v) = -\frac{1}{m(x)} \left( \frac{m'(x)}{2} v^2 - \frac{k_{\mathrm{B}}T m'(x)}{2 m(x)}\right),
\end{equation}
To evaluate the effect the correction term has on the embedding,
we compare the correction term $c(x, v)$ to the difference between $\dv{x} U_{\text{eff}*}(x)$
and $\dv{x} U_{\text{eff}}(x, v)$. We find that the difference between the two potentials is exactly
equal to $c(x, v)$, i.e.
\begin{equation}
    \frac{1}{m(x)} \dv{x} \left[ U_{\text{eff}}(x, v) - U_{\text{eff}*}(x) \right] = -c(x, v).
\end{equation}
We conclude that the embedding in Eq.~\ref{eq:eoms_si_ueff1}
is the same as the embedding in Eq.~\ref{eq:eoms_si}.
Thus, the embedding in Eq.~\ref{eq:eoms_si} also works for the GLE with the effective potential $U_{\text{eff}*}(x)$ in Eq.~\ref{eq:gle_ueff1}.

\FloatBarrier
\section{\label{si:compare}Comparison with Previously Extracted Memory Kernels}
\begin{figure}
    \includegraphics[width=\linewidth]{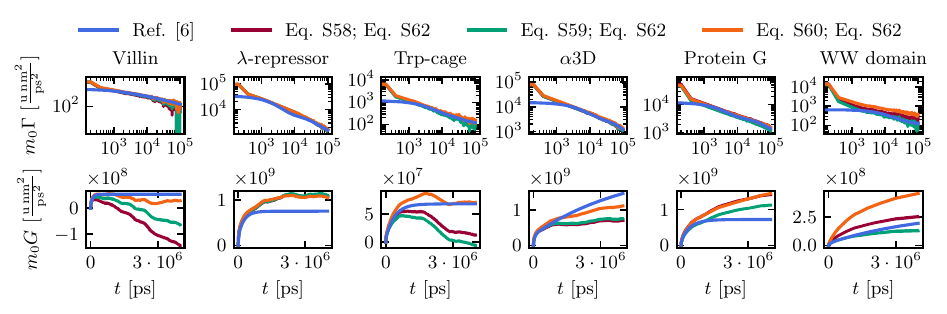}
    \caption{We show kernels based on the approximate GLE in Eq.~\ref{eq:GLE} using the Volterra Eqs.~\ref{eq:volterra1}-\ref{eq:volterra3}, the
 gradient in Eq.~\ref{eq:gradient}, and compare to fits from Ref.~\cite{daltonFastProteinFolding2023}, which used Eq.~\ref{eq:volterra1} and Eq.~\ref{eq:gradient}.}
    \label{fig:compare_kernels1}
\end{figure}

\begin{figure}
    \includegraphics[width=\linewidth]{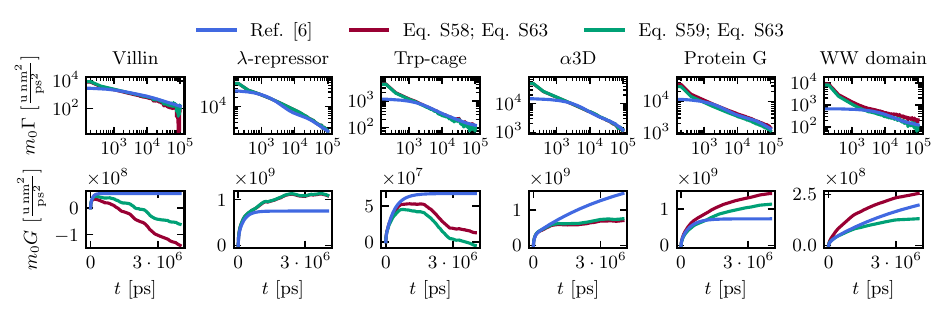}
    \caption{We show kernels based on the approximate GLE in Eq.~\ref{eq:GLE} using the Volterra Eqs. \ref{eq:volterra1} and \ref{eq:volterra2}, the
 gradient in Eq.~\ref{eq:gradient_halfstep}, and compare to fits from Ref.~\cite{daltonFastProteinFolding2023}, which used Eq.~\ref{eq:volterra1} and Eq.~\ref{eq:gradient}.}
    \label{fig:compare_kernels2}
\end{figure}

\begin{figure}
    \includegraphics[scale=0.93]{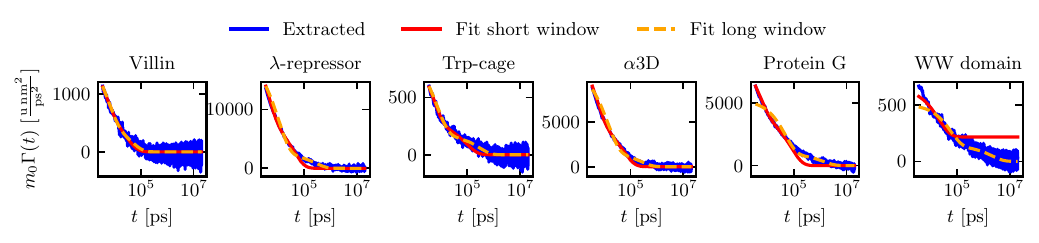}
        \caption{We show kernels based on the approximate GLE in Eq.~\ref{eq:GLE} (blue lines) using the Volterra Eq. \ref{eq:volterra1}, the
 gradient in Eq.~\ref{eq:gradient}, and compare to fits over a short fitting window of 40 ns (red lines) and a long fitting window of 40 $\mu$s (dashed orange lines).
 We have multiplied the kernels by $m_0$ for comparability with Ref.~\cite{daltonFastProteinFolding2023}.
 }
    \label{fig:windows_upmf}
\end{figure}

\begin{figure}
    \includegraphics[scale=0.93]{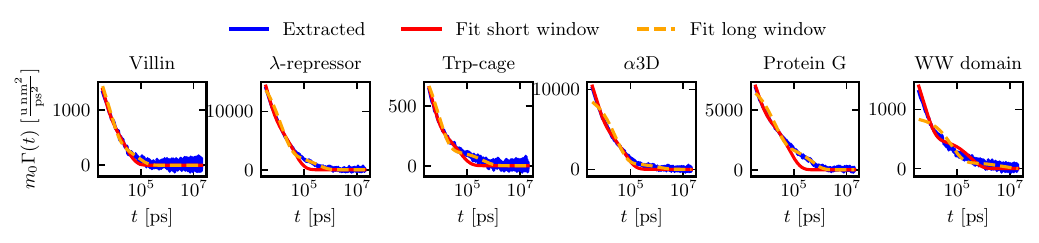}
        \caption{We show kernels based on the GLE in Eq.~\ref{eq:gam_of_x_GLE} (blue lines) using the Volterra Eq. \ref{eq:volterra4}, the
 gradient in Eq.~\ref{eq:gradient}, and compare to fits over a short fitting window of 40 ns (red lines) and a long fitting window of 40 $\mu$s (dashed orange lines).
 We have multiplied the kernels by $m_0$ for comparability with Ref.~\cite{daltonFastProteinFolding2023}.
 }
    \label{fig:windows_ueff2}
\end{figure}

To extract coordinate-independent memory kernels from time series data, several methods exist.
Previous extractions employed the approximate GLE,
including the constant mass term $m_0$, which takes the form
\begin{equation}
 m_0 \dot{v}(t) = -U'_{\text{pmf}}(x(t)) - \int_0^t \dd s \: m_0 \Gamma(t - s) v(s) + m_0 F_R(t),
\end{equation}
which is a rearranged version of the GLE in Eq.~\ref{eq:GLE}. To derive a Volterra equation, we multiply by
$v(0)$ and take an equilibrium average.
Using the fact that the initial velocity is orthogonal to $F_R(t)$, i.e., $m_0 \langle v(0) F_R(t) \rangle = 0$~\cite{ayazNonMarkovianModelingProtein2021},
we obtain
\begin{equation}
 m_0 \dv{t} \langle v(0) v(t) \rangle = -\langle v(0) U'_{\text{pmf}}(x(t)) \rangle -\int_{0}^{t} \dd s \: m_0 \langle v(0) \Gamma\left(t - s\right) v(s) \rangle,
    \label{eq:derive_si}
\end{equation}
which we write as
\begin{equation}
 m_0 \dv{t} C^{vv}(t) = -C^{v U_{\text{pmf}}'}(t) -\int_{0}^{t} \dd s \: m_0 \Gamma\left(t - s\right) C^{vv}(s),
    \label{eq:derive_si1}
\end{equation}
where we have defined the correlation functions $C^{vv}(t) = \langle v(0) v(t) \rangle$ and $C^{v U'_{\text{pmf}}}(t) = \langle v(0) U'_{\text{pmf}}(x(t)) \rangle$.
As before, we integrate Eq.~\ref{eq:derive_si1} and obtain
\begin{equation}
 m_0 \left(C^{vv}(t) - C^{vv}(0) \right) = -\int_0^t \dd s \: C^{v U'_{\text{pmf}}}(s) -\int_0^t \dd s \: C^{vv}(s) m_0 G(t - s)
    \label{eq:derive_si_gmethod}
\end{equation}
with $G(t) = \int_0^t \dd s \: \Gamma(s)$. In contrast to the derivation in the main text for the
GLE in Eq.~\ref{eq:nonlinGLE}, we do not have conditional correlation functions here, which allows us to further
rearrange Eq.~\ref{eq:derive_si_gmethod} as
\begin{equation}
 m_0 \left(C^{vv}(t) - C^{vv}(0) \right) = C^{x U'_{\text{pmf}}}(t) - C^{x U'_{\text{pmf}}}(0) - \int_0^t \dd s \: C^{vv}(s) m_0 G(t - s),
    \label{eq:derive_si_gmethod1}
\end{equation}
where we have defined the correlation function $C^{x U'_{\text{pmf}}}(t) = \langle x(0) U'_{\text{pmf}}(x(t)) \rangle$.
To arrive at a different Volterra equation, in the same way as above,
we multiply Eq.~\ref{eq:GLE} by $x(t)$ and ensemble average, using $\langle x(0) F_R(t) \rangle = 0$~\cite{ayazNonMarkovianModelingProtein2021}, obtaining
\begin{equation}
 m_0 \dv{t} C^{xv}(t) =  -m_0 C^{vv}(t) = -C^{x U'_{\text{pmf}}}(t) - \int_0^t \dd s \: C^{vv}(s) m_0  \Gamma(t - s).
    \label{eq:derive_si_gmethod11}
\end{equation}
Evaluating Eq.~\ref{eq:derive_si_gmethod11} at $t=0$ yields
\begin{equation}
 m_0 = \frac{C^{x U'_{\text{pmf}}}(0)}{C^{vv}(0)} .
    \label{eq:derive_si_gmethod12}
\end{equation}
Inserting Eq.~\ref{eq:derive_si_gmethod12} in Eq.~\ref{eq:derive_si_gmethod1} yields
\begin{equation}
 \frac{C^{x U'_{\text{pmf}}}(0)}{C^{vv}(0)} \left(C^{vv}(t) - C^{vv}(0) \right) = C^{x U'_{\text{pmf}}}(t) - C^{x U'_{\text{pmf}}}(0) - \int_0^t \dd s \: C^{vv}(s) m_0 G(t - s).
    \label{eq:derive_si_gmethod2}
\end{equation}
To obtain the final Volterra-based kernel extraction method, we discretize all functions of time as
$f(t) = f(i \Delta t) = f_i$, discretize the integrals using the trapezoidal rule, and rearrange for
$G(t) = G(i \Delta t) = G_i$, obtaining, based on Eq.~\ref{eq:derive_si_gmethod2}
\begin{equation}
 m_0 G_i = \frac{2 }{\Delta t C^{vv}_0} \left( C^{xU'_{\text{pmf}}}_i - \frac{C^{xU'_{\text{pmf}}}_0}{C^{vv}_0} C^{vv}_i - \Delta t \sum_{j=1}^{i - 1} m_0 G_j C^{vv}_{i - j} \right),
 \label{eq:volterra1}
\end{equation}
based on Eq.~\ref{eq:derive_si_gmethod1}
\begin{equation}
 m_0 G_i = \frac{2 }{\Delta t C^{vv}_0} \left( C^{xU'_{\text{pmf}}}_i - C^{xU'_{\text{pmf}}}_0 + m_0 \left( C^{vv}_0 - C^{vv}_i \right) - \Delta t \sum_{j=1}^{i - 1} m_0 G_j C^{vv}_{i - j} \right),
 \label{eq:volterra2}
\end{equation}
and based on Eq.~\ref{eq:derive_si_gmethod}
\begin{equation}
 m_0 G_i = \frac{2 }{\Delta t C^{vv}_0} \left(\frac{\Delta t}{2} \left( C^{vU'_{\text{pmf}}}_0 - C^{vU'_{\text{pmf}}}_i \right) - \Delta t \sum_{j=1}^{i - 1} C^{vU'_{\text{pmf}}}_j + m_0 \left( C^{vv}_0 - C^{vv}_i \right) - \Delta t \sum_{j=1}^{i - 1} m_0 G_j C^{vv}_{i - j} \right).
 \label{eq:volterra3}
\end{equation}
When using Eq.~\ref{eq:volterra3} to compute kernels based on the GLE in Eq.~\ref{eq:gam_of_x_GLE} in the main text,
we divide by $m_0$, replace $m_0$ by $m(x)$ and $U'_{\text{pmf}}(x)$ by $U'_{\text{eff}}(x, v)$, yielding
\begin{equation}
 G_i = \frac{2 }{\Delta t C^{vv}_0} \left(\frac{\Delta t}{2} \left( C^{vF}_i - C^{vF}_0 \right) - \Delta t \sum_{j=1}^{i - 1} C^{vF}_j + \left( C^{vv}_0 - C^{vv}_i \right) - \Delta t \sum_{j=1}^{i - 1} G_j C^{vv}_{i - j} \right),
 \label{eq:volterra4}
\end{equation}
where we have defined $C^{vF}_i = C^{vF}(i \Delta t) = -\langle v(0) U'_{\text{eff}}(x(t), v(t)) / m(x(t)) \rangle$.\\
To compute the kernel based on Eqs.~\ref{eq:volterra1}-\ref{eq:volterra3}, we need to compute
the velocities $v(t) = v_i$ from the positions $x(t) = x_i$ via numerical gradients. We compare two
methods
\begin{equation}
 v_i = \frac{x_{i + 1} - x_{i - 1}}{2 \Delta t},
    \label{eq:gradient}
\end{equation}
and
\begin{equation}
 v_{i + 1/2} = \frac{x_{i + 1} - x_{i}}{\Delta t}.
    \label{eq:gradient_halfstep}
\end{equation}
We note that velocities computed with Eq.~\ref{eq:gradient_halfstep} are shifted in time with
respect to $x_i$ by $\Delta t / 2$, which means that they cannot be used to compute $C^{vU'}(t)$ in
Eq.~\ref{eq:volterra3} and $C^{vF}(t)$ in Eq.~\ref{eq:volterra4}, and we use the velocities from Eq.~\ref{eq:gradient}
instead. In Figure~\ref{fig:compare_kernels1} and Figure~\ref{fig:compare_kernels2}, we show the
resulting kernels $m_0 \Gamma(t)$ and running integrals over the kernels $m_0 G(t)$ for the three
Volterra equations \ref{eq:volterra1}-\ref{eq:volterra3} and numerical gradients for the velocities
in Eqs.~\ref{eq:gradient} and \ref{eq:gradient_halfstep}. The results show some scatter in $m_0
G(t)$, but $m_0 \Gamma(t)$ is rather similar for all methods. We also plot results from
Ref.~\cite{daltonFastProteinFolding2023}, as a comparison, which is based on the same data, and uses
Eq.~\ref{eq:volterra1} in combination with Eq.~\ref{eq:gradient}. To enable this comparison, all
kernels are shown in the units of Ref.~\cite{daltonFastProteinFolding2023}, i.e., $\frac{\text{u} \,
\text{nm}^2}{\text{ps}^2}$. We see that overall, the data from
Ref.~\cite{daltonFastProteinFolding2023} agrees with our results for $m_0 \Gamma(t)$. \\
The fits in Ref.~\cite{daltonFastProteinFolding2023} use a fitting window of 40 $\mu$s. However,
for our data based on the GLE in Eq.~\ref{eq:nonlinGLE}, we only have on the order of 40 ns available
for the fits due to the increased noise of the data for the conditional correlation functions.
To estimate the effect such a reduced fitting window has, we conduct one set of fits based on the GLEs
in Eq.~\ref{eq:GLE} and Eq.~\ref{eq:gam_of_x_GLE}, where the full 40 $\mu$s are available, and one
where we restrict the fitting window to 40 ns. We show the fits in
Figures~\ref{fig:windows_upmf}-\ref{fig:windows_ueff2} and in Table~\ref{tab:fitting_windows}. We
see that the length of the fitting window has a small effect on the extracted memory kernel for all proteins when using
the GLE in Eq.~\ref{eq:gam_of_x_GLE} and has a small effect for all but one case (which is WW domain) for the GLE in Eq.~\ref{eq:GLE}.
We conclude that fitting over a window length on the order of nanoseconds is sufficient to yield accurate results for the memory kernel.


\begin{table}[h]
 {\scriptsize
    \centering
    \begin{tabular}{|c|c|c|c|c|c|c|c|c|c|}
        \hline
 protein & fit window & GLE & $\tau_1$ [ps] & $\gamma_1$ [$\frac{\text{u}\, \text{nm}^2}{\text{ps}}$] & $\tau_2$ [ps] & $\gamma_2$ [$\frac{\text{u}\, \text{nm}^2}{\text{ps}}$] & $\tau_3$ [ps] & $\gamma_3$ [$\frac{\text{u}\, \text{nm}^2}{\text{ps}}$] & $\gamma_{\mathrm{sum}}$ [$\frac{\text{u}\, \text{nm}^2}{\text{ps}}$] \\ \hline
 Villin               & 40 ns     & Eq.1  & 2.79e+04 & 1.98e+07 & 4.33e+03 & 4.92e+06 & 2.77e+02 & 4.69e+06 & 2.94e+07 \\
 Villin               & 40 ns     & Eq.2  & 2.36e+04 & 2.45e+07 & 2.68e+03 & 5.05e+06 & 3.01e+02 & 3.51e+06 & 3.30e+07 \\
 Villin               & 40 $\mu$s & Eq.1  & 4.48e+04 & 2.17e+07 & 5.63e+03 & 7.28e+06 & 2.79e+02 & 4.74e+06 & 3.37e+07 \\
 Villin               & 40 $\mu$s & Eq.2  & 7.06e+04 & 2.95e+07 & 6.11e+03 & 1.15e+07 & 3.28e+02 & 4.02e+06 & 4.50e+07 \\
 Villin               & Ref.~\cite{daltonFastProteinFolding2023}  & Eq.1  & 9.20e+04 & 4.08e+07 & 9.50e+03 & 1.05e+07 & 1.40e+03 & 2.20e+06 & 5.35e+07 \\ \hline
        $\lambda$-repressor  & 40 ns     & Eq.1  & 3.75e+04 & 3.26e+08 & 3.51e+03 & 5.92e+07 & 2.70e+02 & 3.30e+07 & 4.18e+08 \\
        $\lambda$-repressor  & 40 ns     & Eq.2  & 3.68e+04 & 3.60e+08 & 3.10e+03 & 5.42e+07 & 3.16e+02 & 1.72e+07 & 4.32e+08 \\
        $\lambda$-repressor  & 40 $\mu$s & Eq.1  & 2.68e+05 & 6.57e+08 & 1.08e+04 & 1.73e+08 & 3.04e+02 & 3.89e+07 & 8.69e+08 \\
        $\lambda$-repressor  & 40 $\mu$s & Eq.2  & 3.00e+05 & 7.68e+08 & 1.32e+04 & 1.93e+08 & 4.61e+02 & 2.88e+07 & 9.90e+08 \\
        $\lambda$-repressor  & Ref.~\cite{daltonFastProteinFolding2023}  & Eq.1  & 2.60e+05 & 5.42e+08 & 2.80e+04 & 1.48e+08 & 2.20e+03 & 6.03e+07 & 7.51e+08 \\ \hline
 Trp-cage             & 40 ns     & Eq.1  & 9.14e+04 & 2.45e+07 & 4.68e+03 & 3.36e+06 & 2.81e+02 & 2.43e+06 & 3.03e+07 \\
 Trp-cage             & 40 ns     & Eq.2  & 4.73e+04 & 1.58e+07 & 3.44e+03 & 3.35e+06 & 2.94e+02 & 1.81e+06 & 2.10e+07 \\
 Trp-cage             & 40 $\mu$s & Eq.1  & 2.83e+05 & 4.82e+07 & 7.37e+03 & 5.15e+06 & 2.88e+02 & 2.52e+06 & 5.59e+07 \\
 Trp-cage             & 40 $\mu$s & Eq.2  & 3.81e+05 & 4.75e+07 & 7.98e+03 & 6.77e+06 & 3.26e+02 & 2.11e+06 & 5.63e+07 \\
 Trp-cage             & Ref.~\cite{daltonFastProteinFolding2023}  & Eq.1  & 3.90e+05 & 6.09e+07 & 1.70e+04 & 4.80e+06 & 2.90e+03 & 2.22e+06 & 6.79e+07 \\ \hline
        $\alpha$3D           & 40 ns     & Eq.1  & 5.22e+04 & 2.63e+08 & 4.48e+03 & 4.37e+07 & 2.74e+02 & 2.89e+07 & 3.35e+08 \\
        $\alpha$3D           & 40 ns     & Eq.2  & 5.44e+04 & 3.27e+08 & 4.05e+03 & 4.73e+07 & 3.32e+02 & 2.04e+07 & 3.94e+08 \\
        $\alpha$3D           & 40 $\mu$s & Eq.1  & 2.50e+05 & 3.99e+08 & 1.40e+04 & 1.29e+08 & 2.97e+02 & 3.24e+07 & 5.60e+08 \\
        $\alpha$3D           & 40 $\mu$s & Eq.2  & 1.10e+06 & 8.44e+08 & 2.82e+04 & 2.45e+08 & 4.68e+02 & 3.07e+07 & 1.12e+09 \\
        $\alpha$3D           & Ref.~\cite{daltonFastProteinFolding2023}  & Eq.1  & 4.30e+06 & 1.96e+09 & 6.30e+04 & 2.39e+08 & 4.90e+03 & 5.36e+07 & 2.25e+09 \\ \hline
 Protein G            & 40 ns     & Eq.1  & 7.85e+04 & 3.17e+08 & 5.01e+03 & 2.55e+07 & 2.88e+02 & 1.93e+07 & 3.62e+08 \\
 Protein G            & 40 ns     & Eq.2  & 6.43e+04 & 3.13e+08 & 3.28e+03 & 2.38e+07 & 3.04e+02 & 1.21e+07 & 3.49e+08 \\
 Protein G            & 40 $\mu$s & Eq.1  & 1.35e+06 & 1.22e+09 & 4.81e+04 & 2.08e+08 & 3.31e+02 & 2.33e+07 & 1.45e+09 \\
 Protein G            & 40 $\mu$s & Eq.2  & 4.53e+05 & 9.13e+08 & 1.91e+04 & 1.02e+08 & 3.96e+02 & 1.72e+07 & 1.03e+09 \\
 Protein G            & Ref.~\cite{daltonFastProteinFolding2023}  & Eq.1  & 4.10e+05 & 5.88e+08 & 4.30e+04 & 1.14e+08 & 2.80e+03 & 2.24e+07 & 7.24e+08 \\ \hline
 WW domain            & 40 ns     & Eq.1  & 1.32e+14 & 2.84e+16 & 2.07e+04 & 8.83e+06 & 3.18e+02 & 5.21e+06 & 2.84e+16 \\
 WW domain            & 40 ns     & Eq.2  & 3.25e+05 & 1.63e+08 & 5.51e+03 & 9.50e+06 & 2.95e+02 & 5.12e+06 & 1.77e+08 \\
 WW domain            & 40 $\mu$s & Eq.1  & 2.13e+06 & 2.86e+08 & 5.89e+04 & 2.18e+07 & 3.24e+02 & 5.36e+06 & 3.13e+08 \\
 WW domain            & 40 $\mu$s & Eq.2  & 5.07e+06 & 5.10e+08 & 6.51e+04 & 5.02e+07 & 3.65e+02 & 6.68e+06 & 5.67e+08 \\
 WW domain            & Ref.~\cite{daltonFastProteinFolding2023}  & Eq.1  & 4.50e+06 & 2.98e+08 & 1.30e+05 & 1.61e+07 & 1.06e+04 & 5.10e+06 & 3.19e+08 \\ \hline
        \hline
    \end{tabular}
 } \caption{ Parameters of the multi-exponential fits function in Eq.~\ref{eq:kernel_ben} with three
    components based on the GLEs in Eqs.~\ref{eq:GLE} and \ref{eq:gam_of_x_GLE}, for a short fitting
    window (40 ns) and a long window (40 $\mu$s). We have multiplied the kernels by $m_0$ for
    comparability with the parameters from Ref.~\cite{daltonFastProteinFolding2023}, which is also listed. }
 \label{tab:fitting_windows}
\end{table}

\FloatBarrier
\section{\label{si:pos_indep_kernel}Memory Kernel Extraction for the GLE in Eq.~\ref{eq:gam_of_x_GLE}}
We extract the kernel for the GLE in Eq.~\ref{eq:gam_of_x_GLE} in the main text from the MD data using the RC-independent analog of
the Volterra Eq.~\ref{eq:derive_final}
\begin{equation}
 C^{vv}(t) - C^{vv}(0) = \int_0^t \dd s \: C^{vF}(s) -\int_0^t \dd s \: C^{vv}(s) G(t - s).
 \label{eq:volterra_new_gle}
\end{equation}
As in Section~\ref{si:numerics_volterra}, we compute the coordinate-independent time correlation functions $C^{ab}(t) = \langle a(0)  b(t)
\rangle$ via a discrete version of the Wiener-Khinchin
theorem~\cite{wienerGeneralizedHarmonicAnalysis1930,khintchineKorrelationstheorieStationaerenStochastischen1934}
\begin{equation}
 C^{ab}_i = \frac{1}{N - i} IDFT[DFT[a_i] \odot DFT[b_i]^*],
\end{equation}
where $a_i = a(i \Delta t)$ and $b_i = b(i \Delta t)$ are two discrete time series of length $N$,
which are zero-padded by appending $N$ zeros, where $N$ is the length of $a$ and $b$.
We use the 'Powell' method of \texttt{scipy.optimize.minimize}~\cite{2020SciPy-NMeth} to minimize the mean-squared loss
\begin{equation}
    \mathcal{L}(\tau_i, \gamma_i) = \frac{1}{N} \sum_{t = 1}^N
 \left( \Gamma(t) - \sum_{i=1}^{2} \frac{\gamma_i}{\tau_i} e^{-t / \tau_i}\right)^2.
    \label{eq:loss_no_pos}
\end{equation}
For the GLE simulations, we use the embedding in Eq.~\ref{eq:eoms_si}, which produces the kernel in
Eq.~\ref{eq:kernel_sim}. We note that in order to obtain an embedding for a kernel $\Gamma(t) =
\sum_{i=1}^2 \gamma_i / \tau_i \exp\{ -t / \tau_i \}$ that does not depend on the position, we
set $\bar{\gamma}_i(x(t))$ in Eq.~\ref{eq:eoms_si} to $\bar{\gamma}_i(x(t)) = \gamma_i / m(x(t))$, with the
parameters $\gamma_i$ and $\tau_i$ listed in Table~\ref{tab:fitting_windows} under 'GLE: Eq.2' and
'fit window: 40 $\mu$s'.

\begin{figure}
    \centering
    \includegraphics[trim={0.2cm 0 0 0 },clip]{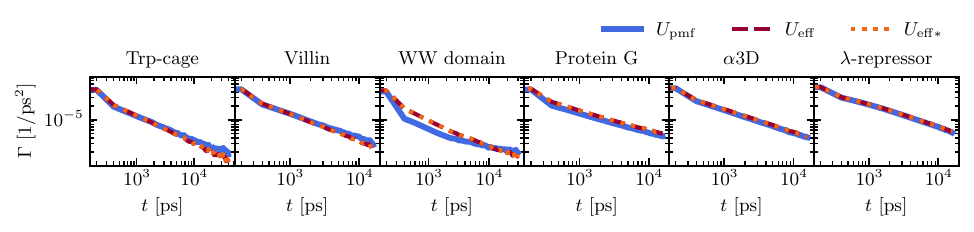} 
    \caption{We compare the kernels using $U_{\text{pmf}}(x)$, based on the GLE in Eq.~\ref{eq:GLE}
 (blue lines), kernels based on the GLE in Eq.~\ref{eq:gam_of_x_GLE} using the potentials
 $U_{\text{eff}}(x, v)$ in Eq.~\ref{eq:ueff} (dashed red lines) and $U_{\text{eff}*}(x)$ in
 Eq.~\ref{eq:gle_ueff1}, (dotted orange lines).
}%
    \label{figsi:ker_ueff1}%
\end{figure}

Figure~\ref{figsi:ker_ueff1} shows that kernels using $U_{\text{eff}}(x, v)$ (dark red dashed lines) in
Eq.~\ref{eq:ueff} and $U_{\text{eff}*}(x)$ (orange dotted lines) in Eq.~\ref{eq:gle_ueff1}, based on the GLE
in Eq.~\ref{eq:gam_of_x_GLE}, are almost identical and very similar to kernels based on
$U_{\text{pmf}}(x)$ (blue lines).


\FloatBarrier

\end{document}